\documentclass[AMA,STIX1COL]{WileyNJD-v2}

\articletype{Research Article}%

\usepackage{subcaption}
\usepackage[font=normalsize,singlelinecheck=off,justification=raggedright]{subcaption}
\usepackage[nameinlink,noabbrev,capitalize]{cleveref} 
\usepackage{bm}
\usepackage{lineno}
\usepackage{orcidlink}
\usepackage{amsmath}
\usepackage{anyfontsize}
\hypersetup{colorlinks=true,allcolors=blue}

\begin{document}

\title{NExT-LF: A Novel Operational Modal Analysis Method via Tangential Interpolation\protect\thanks{\hl{This is the peer reviewed version of the following article: G. Dessena, M. Civera, A. Yousefi, and C. Surace, “NExT-LF: A Novel Operational Modal Analysis Method via Tangential Interpolation”, International Journal of Mechanical System Dynamics. Wiley, May 13, 2025. doi: 10.1002/msd2.70016, which has been published in final form and open access at} [\url{https://doi.org/10.1002/msd2.70016}]. \hl{This is an open access article distributed under the terms of the Creative Commons Attribution License (CC BY 4.0), which permits use, distribution and reproduction in any medium, provided the original work is properly cited. To view a copy of this license, visit} \url{http://creativecommons.org/licenses/by/4.0/}\hl{. The article must be cited appropriately and linked to Wiley’s version of record on Wiley Online Library.}}}

\author[1]{Gabriele Dessena}

\author[2]{Marco Civera}

\author[2]{Ali Yousefi}

\author[2]{Cecilia Surace}

\authormark{Gabriele Dessena \textsc{et al}}

\address[1]{\orgdiv{Department of Aerospace Engineering}, \orgname{Universidad Carlos III de Madrid}, \orgaddress{\state{Leganés, Madrid}, \country{Spain}}}

\address[2]{\orgdiv{Department of Structural, Geotechnical and Building Engineering}, \orgname{Politecnico di Torino}, \orgaddress{\state{Turin}, \country{Italy}}}

\corres{Dr Marco Civera, Department of Structural, Geotechnical and Building Engineering, Politecnico di Torino, Corso Duca degli Abruzzi 24, 10129, Turin, Italy. \\
\email{\href{mailto:marco.civera@polito.it}{marco.civera@polito.it}}}

\fundingInfo{Madrid Government (\emph{Comunidad de Madrid} -- Spain) under the Multiannual Agreement with the Universidad Carlos III de Madrid (UC3M) (IA\_aCTRl-CM-UC3M); Grants for research activity of young PhD holders, part of the UC3M Own Research Program ({\emph{Ayudas para la Actividad Investigadora de los Jóvenes Doctores, del Programa Propio de Investigación de la UC3M}}); \emph{Centro Nazionale per la Mobilità Sostenibile} (MOST - Sustainable Mobility Center), Spoke 7 (Cooperative Connected and Automated Mobility and Smart Infrastructures), Work Package 4 (Resilience of Networks, Structural Health Monitoring and Asset Management).
}

\abstract[Abstract]{Operational Modal Analysis (OMA) is vital for identifying modal parameters under real-world conditions, yet existing methods often face challenges with noise sensitivity and stability. This work introduces NExT-LF, a novel method that combines the well-known Natural Excitation Technique (NExT) with the Loewner Framework (LF). NExT enables the extraction of Impulse Response Functions (IRFs) from output-only vibration data, which are then converted into the frequency domain and used by LF to estimate modal parameters. The proposed method is validated through numerical and experimental case studies. In the numerical study of a 2D Euler-Bernoulli cantilever beam, NExT-LF provides results consistent with analytical solutions and {those from standard methods,} NExT with Eigensystem Realization Algorithm (NExT-ERA) {and Stochastic Subspace Identification with Canonical Variate Analysis (SSI)}. Additionally, NExT-LF demonstrates superior noise robustness, reliably identifying stable modes across various noise levels where NExT-ERA fails. Experimental validation on the Sheraton Universal Hotel is the first OMA application to this structure, confirming NExT-LF as a robust and efficient method for output-only modal parameter identification.}


\jnlcitation{\cname{%
\author{G. Dessena},
\author{M. Civera}, 
\author{A. Yousefi}, and 
\author{C. Surace}} (\cyear{2025}), 
\ctitle{NExT-LF: A Novel Operational Modal Analysis Method via Tangential Interpolation}, \cjournal{Int. J. Mech. Syst. Dyn.}, doi: \href{https://doi.org/10.1002/msd2.70016}{10.1002/msd2.70016}}

\maketitle

\section{Introduction}\label{sec:intro}

The importance of modal analysis in modern engineering is undisputed. Modal parameters -- namely natural frequencies ($\omega_n$), damping ratios ($\zeta_n$), and mode shapes ($\bm{\phi}_n$) -- are used for a variety of tasks across various engineering systems, such as aircraft wings,\cite{Dessena2022b} fuel pumps,\cite{Verhulst2022} and civil structures.\cite{Civera2021a} Nevertheless, this characterisation is not carried out for its own sake, but {to inform} the design process, such as in aircraft certification,\cite{lubrina2014} updating existing finite element models, \cite{Dessena2022c} and/or for structural health monitoring (SHM). \cite{Cadoret2025,Farrar2025} In particular, modal analysis has been used to monitor the health of infrastructure such as railway bridges \cite{Liu2009} and tall buildings,\cite{Kavyashree2021}, while also revealing the effect of complex fluid-structure interactions on systems dynamics, such as those in conical shells partially filled with fluids.\cite{Bochkarev2024} Advanced techniques have also enabled dynamic parameter identification in robotic systems, overcoming nonlinearities to enhance their control precision.\cite{Li2024}

In general, there are two main approaches for modal analysis: Experimental Modal Analysis (EMA) and Operational Modal Analysis (OMA). EMA relies on controlled excitations to characterise a structure dynamics accurately. Due to the requirement of a controlled environment, EMA is resource-intensive and constrained by the need for specialised setups, making it unsuitable for many large-scale or in-situ applications.\cite{Sibille2023} By contrast, OMA exploits ambient vibrations, eliminating the need for external input forces, such as those generated by shaker tables.\cite{Aglietti2019} This makes OMA particularly advantageous for scenarios such as monitoring long-span bridges \cite{Peeters2015}, evaluating aerospace systems during operation,\cite{Pappa1998} and assessing historic buildings.\cite{Gentile2007}

Central to EMA and OMA is system identification (SI), here defined as the process of extracting dynamic properties and/or models from measured data. \cite{Ljung2020} SI techniques generally fall into two categories: time-domain and frequency-domain methods. Within these, input-output approaches, as used in EMA, rely on known external excitations to establish dynamic relationships (such as frequency response functions -- {FRFs}). Conversely, output-only techniques, such as those employed in OMA, can use stochastic ambient excitations, by modelling them as random processes.\cite{Reynders2012} Recent advances in SI, including artificial intelligence and machine learning techniques, have streamlined key processes, such as interpreting stabilisation diagrams or denoising. However, while these innovations improve automation and accuracy, they often introduce computational complexity and may require higher data quality.\cite{Mugnaini2022} 

Moreover, {new modal identification techniques, as well as improved variants of existing algorithms, have been} recently introduced to address the noise robustness and mode consistency problem. In particular, a robust implementation of the stochastic subspace identification covariance-driven variant (Cov-SSI) has been proposed in \cite{OConnell2024} to incorporate a probabilistic method to reduce {the risk of misidentifying outliers as physical modes}. Furthermore, a dual Unscented Kalman filter approach to simultaneously evaluate changes to the mass distribution and kinematic state of an in-orbit satellite flexible dynamics has been introduced in \cite{A.J.Elliott2025} {and a Bayesian approach for the OMA of systems having closely spaced modes has been presented in }\cite{Zhu2019} {and applied to an existing suspension footbridge.} Notably, a single-input multi-output (SIMO) identification method widely applied in electrical engineering, \cite{Grivet2016} the Fast Relaxed Vector Fitting, has been {successfully applied in modal identification and vibration-based SHM} to exploit its computational efficiency.\cite{Civera2021}

In this wave of emerging SI techniques, the Loewner Framework (LF),\cite{Dessena2022} also coming from electrical engineering,\cite{Lefteriu2009} has distinguished itself as a powerful tool for SIMO system identification, offering strong noise resilience and high accuracy in modal parameters estimation. However, these have not yet been proven in an OMA (unknown input) setting. Nevertheless, the Natural Excitation Technique (NExT) \cite{Carne1988} has demonstrated significant capability in output-only applications, effectively allowing the use of ambient vibrations with input-output modal identification methods.\cite{Lauffer1988} These two methods, although historically separate, present complementary strengths. Hence, this work introduces a novel combination of these techniques, referred to as the NExT-LF approach, to tackle OMA. By combining the computational simplicity of NExT with the noise rejection capabilities of the LF, the proposed method strives to address the limitations, in existing techniques {(such as NExT-ERA)} of noise sensitivity. Hence, the pairing of NExT and the Eigensystem realization algorithm (ERA),\cite{Juang1986} NExT-ERA,\cite{GeorgeIII1993} is used as a benchmark method in this work{, alongside -- for the numerical system-only -- the well-known Stochastic Subspace Identification with Canonical Variate Analysis (SSI) }.\cite{VanOverschee1996} Thus, the main contributions of this work are as follows:
\begin{itemize}
    \item The development of the NExT-LF method, integrating the LF with NExT, to achieve OMA capability;
    \item Numerical validation of NExT-LF, across varying noise conditions on a numerical system;
    \item Experimental validation of the method on a first-in-literature identified structure: The Sheraton Universal Hotel in Universal City, North Hollywood, California (USA).
\end{itemize}

To achieve its aims, this work tackles the following:
\begin{enumerate}
    \item In \cref{sec:met} (Methodology), the LF identification method is introduced, before discussing the output-only enabling method, NExT, and their pairing;
    \item In \cref{sec:num} (Numerical Case Study), the proposed method is applied to a numerical of a cantilever beam to preliminary assess its noise robustness characteristics;
    \item In \cref{sec:exp} (Operational Modal Analysis of the Sheraton Universal Hotel), the Sheraton Universal Hotel environmental vibration is analysed and the modal parameters extracted via NExT-LF and NExT-ERA;
    \item The conclusions (\cref{sec:conc}) end this paper.
\end{enumerate}

\section{Methodology}\label{sec:met}

\subsection{Loewner Framework}\label{sec:lf}

Previously, the LF algorithm has been employed for the modelling of a multi-port electrical system \cite{Lefteriu2009} and utilised for aerodynamic model order reduction in the context of aeroservoelastic modelling.\cite{Quero2019} Later, the first and second authors applied the LF for the identification of modal parameters from SIMO mechanical systems in,\cite{Dessena2022} verified its computational efficiency in, \cite{Dessena2022f} and later assessed its robustness to noise for SHM in.\cite{Dessena2022g} Further developments have been carried out for the extension of the LF for the extraction of modal parameters from multi-input multi-output systems in,\cite{Dessena2024,Dessena2024c} but the version considered in this work is the SIMO version first introduced in.\cite{Dessena2022}

Let us begin by defining the Loewner matrix ${\bm{\mathbb{L}}}$:
\noindent \emph{Given a row array of pairs of complex numbers ($\mu_j$,${v}_j$), $j=1$,...,$q$, and a column array of pairs of complex numbers ($\lambda_i$,${w}_j$), $i=1$,...,$k$, with $\lambda_i$, $\mu_j$ distinct, the associated $\boldsymbol{\bm{\mathbb{L}}}$, or divided-differences matrix is:}
\begin{equation}
\label{eq:LM}
\boldsymbol{\bm{\mathbb{L}}}=\begin{bmatrix}
\frac{\bm{v}_1-\bm{w}_1}{\mu_1-\lambda_1} & \cdots & \frac{\bm{v}_1-\bm{w}_k}{\mu_1-\lambda_k}\\
\vdots & \ddots & \vdots\\
\frac{\bm{v}_q-\bm{w}_1}{\mu_q-\lambda_1} & \cdots & \frac{\bm{v}_q-\bm{w}_k}{\mu_q-\lambda_k}\\
\end{bmatrix}\:\in \mathbb{C}^{q\times k}
\end{equation}
\emph{If there is a known underlying function $\pmb{\phi}$, then $\bm{w}_i=\pmb{\phi}(\lambda_i)$ and $\bm{v}_j=\pmb{\phi}(\mu_j).$}

Karl Löwner established a relationship between $\boldsymbol{\bm{\mathbb{L}}}$ and rational interpolation, often referred to as Cauchy interpolation. \cite{Lowner1934} This connection enables the definition of interpolants through the determinants of submatrices of $\boldsymbol{\bm{\mathbb{L}}}$. As shown in,\cite{Antoulas2017,Mayo2007} rational interpolants can be obtained directly from $\boldsymbol{\bm{\mathbb{L}}}$. This study adopts the approach based on the Loewner pencil, which comprises the $\boldsymbol{\bm{\mathbb{L}}}$ and $\boldsymbol{\bm{\mathbb{L}}}_s$ matrices. Here, $\boldsymbol{\bm{\mathbb{L}}}_s$ represents the \emph{Shifted Loewner matrix}, a concept to be defined later.

To describe the working principle of the LF, consider a linear time-invariant dynamical system $\bm{\Sigma}$ characterised by $m$ inputs, $p$ outputs, and $k$ internal variables, represented in descriptor form as:
\begin{equation}
\bm{\Sigma}:\;\bm{E}\frac{d}{dt}\bm{x}(t)=\bm{A}\bm{x}(t)+\bm{B}\bm{u}(t);\;\;\;
\bm{y}(t)=\bm{C}\bm{x}(t)+\bm{D}\bm{u}(t)
 \label{eq:LTI}
 \end{equation}
\noindent where $\bm{x}(t):\in: \mathbb{R}^{k}$ represents the internal variable, $\bm{u}(t):\in:\mathbb{R}^{m}$ denotes the input function, and $\bm{y}(t):\in:\mathbb{R}^{p}$ corresponds to the output. The constant system matrices are:
\begin{equation}
    \bm{E},\bm{A}\in \mathbb{R}^{k\times k},\; \bm{B}\in \mathbb{R}^{k\times m}\; \bm{C}\in \mathbb{R}^{p\times k}\; \bm{D}\in \mathbb{R}^{p\times m}
\end{equation}
\noindent The Laplace transfer function, $\bm{H}(s)$, of $\bm{\Sigma}$ can be expressed as a $p \times m$ rational matrix function, provided that the matrix $\bm{A}-\lambda\bm{E}$ is non-singular for a given finite value $\lambda$, where $\lambda \in \mathbb{C}$:
\begin{equation}
    \bm{H}(s)=\bm{C}(s\bm{E}-\bm{A})^{-1}\bm{B}+\bm{D}
    \label{eq:trans}
\end{equation}
\noindent Let us examine the general framework of tangential interpolation, commonly identified as rational interpolation along tangential directions.\cite{Kramer2016} The associated right interpolation data is expressed as:
\begin{equation}
\begin{gathered}
   (\lambda_i;\bm{r}_i,\bm{w}_i),\: i = 1,\dots,\rho
    \\
    \begin{matrix}
        \bm{\Lambda}=\text{diag}[\lambda_1,\dotsc,\lambda_k]\in \mathbb{C}^{\rho\times \rho}\\
        \bm{R}=[\bm{r}_1\;\dotsc \bm{r}_k]\in \mathbb{C}^{m\times \rho}\\
        \bm{W} = [\bm{w}_1\;\dotsc\;\bm{w}_k]\in \mathbb{C}^{\rho\times \rho}
        \end{matrix}\Bigg\}
\end{gathered}
\label{eq:RID}
\end{equation}
\noindent Similarly, the left interpolation data is defined as:
\begin{equation}
    \begin{gathered} 
        (\mu_j,\bm{l}_j,\bm{v}_j),\: j = 1,\dots,v
        \\ 
        \begin{matrix}
            \bm{M}=\mathrm{diag}[\mu_1,\dotsc,\mu_q]\in \mathbb{C}^{v\times v}\\
            \bm{L}^\text{T}=[\bm{l}_1\;\dotsc \bm{l}_v]\in \mathbb{C}^{p\times v}\\
            \bm{V}^\text{T} = [\bm{v}_1\;\dotsc\;\bm{v}_q]\in \mathbb{C}^{m\times v}
        \end{matrix}\Bigg\}
    \end{gathered}
\label{eq:LID}
\end{equation}
\noindent The values $\lambda_i$ and $\mu_j$ correspond to the points at which $\bm{H}(s)$ is evaluated, representing the frequency bins in this context. The vectors $\bm{r}_i$ and $\bm{l}_j$ denote the right and left tangential general directions, which are typically chosen randomly in practice,\cite{Quero2019} while $\bm{w}_i$ and $\bm{v}_j$ represent the respective tangential data. Establishing a connection between $\bm{w}_i$ and $\bm{v}_j$ and the transfer function $\bm{H}$, linked to the realisation $\bm{\Sigma}$ in \cref{eq:LTI}, resolves the rational interpolation problem:
\begin{equation}
 \begin{split}
\bm{H}(\lambda_i)\bm{r}_i=\bm{w}_i,\:j=1.\dots,\rho \\ \text{and} \\
\bm{l}_i\bm{H}(\mu_j)=\bm{v}_j,\:i=1,\dots,v
 \end{split}
 \label{eq:LS2}
 \end{equation}
\noindent ensuring that the Loewner pencil satisfies \cref{eq:LS2}.
Next, consider a set of points $Z = {z_1, \dots, z_N}$ in the complex plane and a rational function $\bm{y}(s)$, where $\bm{y}_i := \bm{y}(z_i)$ for $i = 1, \dots, N$, with $Y = {\bm{y}_1, \dots, \bm{y}_N}$. By incorporating the left and right data partitions, the following expressions are obtained:
\begin{equation}
\begin{split}
	Z=\{\lambda_1,\dots,\lambda_\rho\} \cup \{\mu_1,\dots,\mu_v\}\\ \text{and} \\
	Y=\{\bm{w}_1,\dots,\bm{w}_\rho\} \cup \{\bm{v}_1,\dots,\bm{v}_v\}
\end{split}
\label{eq:ZY}
\end{equation}
\noindent where $N = p + v$. Consequently, the matrix $\boldsymbol{\bm{\mathbb{L}}}$ is presented as:
\begin{equation}
\label{eq:LM2}
\boldsymbol{\bm{\mathbb{L}}}=\begin{bmatrix}
\frac{\bm{v}_1\bm{r}_1-\bm{l}_1\bm{w}_1}{\mu_1-\lambda_1} & \cdots & \frac{\bm{v}_1\bm{r}\rho-\bm{l}_1\bm{w}\rho}{\mu_1-\lambda\rho}\\
\vdots & \ddots & \vdots\\
\frac{\bm{v}_v\bm{r}_1-\bm{l}_v\bm{w}_1}{\mu_v-\lambda_1}& \cdots & \frac{\bm{v}_v\bm{r}\rho-\bm{l}_v\bm{w}\rho}{\mu_v-\lambda\rho}\\
\end{bmatrix}\:\in \mathbb{C}^{v\times \rho}
\end{equation}

\noindent As $\bm{v}_v\bm{r}_p$ and $\bm{l}_v\bm{w}_p$ are scalars, the Sylvester equation for $\boldsymbol{\bm{\mathbb{L}}}$ is satisfied in the following manner:
\begin{equation}
    \boldsymbol{\bm{\mathbb{L}}}\bm{\Lambda}-\bm{M}\boldsymbol{\bm{\mathbb{L}}}=\bm{L}\bm{W}-\bm{V}\bm{R}
    \label{eq:syl}
\end{equation}

\noindent The \emph{shifted Loewner matrix}, $\boldsymbol{\bm{\mathbb{L}}}_s$, is defined as the matrix $\boldsymbol{\bm{\mathbb{L}}}$ corresponding to $s\bm{H}(s)$:

\begin{equation}
\label{eq:LS}
\boldsymbol{\bm{\mathbb{L}}}_s=\begin{bmatrix}
\frac{\mu_1\bm{v}_1\bm{r}_1-\lambda_1\bm{l}_1\bm{w}_1}{\mu_1-\lambda_1} & \cdots & \frac{\mu_1\bm{v}_1\bm{r}_\rho-\lambda_\rho\bm{l}_1\bm{w}_\rho}{\mu_1-\lambda_\rho}\\
\vdots & \ddots & \vdots\\
\frac{\mu_v\bm{v}_v\bm{r}_1-\lambda_1\bm{l}_v\bm{w}_1}{\mu_v-\lambda_1}& \cdots & \frac{\bm{v}_v\bm{r}_\rho-\bm{l}_v\bm{w}_\rho}{\mu_v-\lambda_\rho}\\
\end{bmatrix}\:\in \mathbb{C}^{v\times \rho}
\end{equation}

\noindent Similarly, the Sylvester equation is satisfied as follows:
\begin{equation}
    \boldsymbol{\bm{\mathbb{L}}}_s\Lambda-\bm{M}\boldsymbol{\bm{\mathbb{L}}}_s=\bm{L}\bm{W}\bm{\Lambda}-\bm{M}\bm{V}\bm{R}
    \label{eq:syl2}
\end{equation}

\noindent Without loss of generality, $\bm{D}$ can be considered $0$, as its contribution does not influence the tangential interpolation of the LF.\cite{Mayo2007} For ease of presentation, the remainder of the discussion will focus on $\bm{H}(s)$. Consequently, \cref{eq:trans} simplifies to:

\begin{equation}
\bm{H}(s)=\bm{C}(s\bm{E}-\bm{A})^{-1}\bm{B}
\label{eq:fin}
\end{equation}

A minimal-dimensional realisation is achievable only when the system is fully controllable and observable. Under the assumption that the data is sampled from a system whose transfer function is described by \cref{eq:fin}, the generalised tangential observability, $\mathcal{O}_v$, and generalised tangential controllability, $\mathcal{R}_\rho$, are defined in,\cite{Lefteriu2010b} such that \cref{eq:LM2} and \cref{eq:LS} can be rewritten as:
\begin{align}
    \boldsymbol{\bm{\mathbb{L}}}=-\mathcal{O}_v\bm{E}\mathcal{R}_\rho &&
    \boldsymbol{\bm{\mathbb{L}}}_s=-\mathcal{O}_v\bm{A}\mathcal{R}_\rho
    \label{eq:LLs}
\end{align}
Then, by defining the Loewner pencil as a regular pencil, such that $\mathrm{eig}((\boldsymbol{\bm{\mathbb{L}}},\boldsymbol{\bm{\mathbb{L}}}_s)) \neq (\mu_i, \lambda_i)$:
\begin{align}
    \bm{E}=-\boldsymbol{\bm{\mathbb{L}}},&&\bm{A}=-\boldsymbol{\bm{\mathbb{L}}}_s,&&\bm{B}=\bm{V},&&\bm{C}=\bm{W}
\end{align}
As a result, the interpolating rational function is defined as:
\begin{equation}
    \bm{H}(s)=\bm{W}(\boldsymbol{\bm{\mathbb{L}}}_s-s\boldsymbol{\bm{\mathbb{L}}})^{-1}\bm{V}
\end{equation}
The derivation provided is specific to the minimal data scenario, which is rarely encountered in practical {applications}. However, the LF framework can be extended to handle redundant data points effectively. To begin, assume:
\begin{equation}
\begin{split}
    \mathrm{rank}[\zeta\boldsymbol{\bm{\mathbb{L}}}-\boldsymbol{\bm{\mathbb{L}}}_s]=\mathrm{rank}[\boldsymbol{\bm{\mathbb{L}}}\:\boldsymbol{\bm{\mathbb{L}}}_s]
   =\mathrm{rank}
    \begin{bmatrix}
    \boldsymbol{\bm{\mathbb{L}}} \boldsymbol{\bm{\mathbb{L}}}_s
    \end{bmatrix}=k,\; 
    \forall \zeta \in \{\lambda_j\}\cup\{\mu_i\}
    \end{split}
    \label{eq:cond1}
\end{equation}
Next, the short Singular Value Decomposition (SVD) of $\zeta\boldsymbol{\bm{\mathbb{L}}} - \boldsymbol{\bm{\mathbb{L}}}_s$ is performed:
\begin{equation}
    \textrm{svd}(\zeta\boldsymbol{\bm{\mathbb{L}}}-\boldsymbol{\bm{\mathbb{L}}}_s)=\bm{Y}\bm{\Sigma}_l\bm{X}
\label{eq:cond2}
\end{equation}
where $\mathrm{rank}(\zeta\boldsymbol{\bm{\mathbb{L}}}-\boldsymbol{\bm{\mathbb{L}}}_s)=\mathrm{rank}(\bm{\Sigma}_l)=\mathrm{size}(\bm{\Sigma}_l)=k,\bm{Y}\in\mathbb{C}^{v \times k}$ and $\bm{X}\in\mathbb{C}^{k\times \rho}$.
Observe that:
\begin{equation}
\begin{split}
-\bm{A}\bm{X}+\bm{E}\bm{X}\bm{\Sigma}_l = \bm{Y}^*\boldsymbol{\bm{\mathbb{L}}}_s\bm{X}^*\bm{X}-\bm{Y}^*\boldsymbol{\bm{\mathbb{L}}}\bm{X}^*\bm{X}\bm{\Sigma}_l=\bm{Y}^*(\boldsymbol{\bm{\mathbb{L}}}_s-\boldsymbol{\bm{\mathbb{L}}}\bm{\Sigma}_l )=\bm{Y}^*\bm{V}\bm{R}=\bm{BR}
\end{split}
\label{eq:cond3}
\end{equation}
Similarly, $-\bm{Y}\bm{A} + \bm{M}\bm{Y}\bm{E} = \bm{L}\bm{C}$, where $\bm{X}$ and $\bm{Y}$ represent the generalised controllability and observability matrices, respectively, for the system $\bm{\Sigma}$ with $\bm{D} = 0$.
Once the right and left interpolation conditions are verified, the Loewner realisation for redundant data is expressed as:
\begin{equation}
\begin{split}
\bm{E}=-\bm{Y}^*\boldsymbol{\bm{\mathbb{L}}}\bm{X},\quad \quad
\bm{A}=-\bm{Y}^*\boldsymbol{\bm{\mathbb{L}}}_s\bm{X,} \quad \quad
\bm{B}=\bm{Y}^*\bm{V},\quad \quad
\bm{C}=\bm{W}\bm{X}
\end{split}
\label{eq:real}
\end{equation}

The formulation in \cref{eq:real}, which represents the Loewner realisation for redundant data, will be used throughout this work. For a comprehensive explanation of each step, readers are directed to,\cite{Mayo2007,Antoulas2017} while the MATLAB implementation can be found at.\cite{Dessena2021} Finally, the system modal parameters can be determined through eigenanalysis of the system matrices $\bm{A}$ and $\bm{C}$ in \cref{eq:real}.

\subsection{Natural Excitation Technique}\label{sec:next}
The NExT is a method used to extract impulse response functions (IRF) of structures subjected to ambient excitation.\cite{GeorgeIII1993} Originally applied to wind turbines,\cite{Lauffer1988,Carne1988} this approach is particularly suited to scenarios where input forces, such as wind or traffic, cannot be directly measured. The NExT algorithm works under the principle that ambient excitations serve as random broadband inputs, effectively exciting the structure.

NExT requires {that the ambient vibration responses (in terms of acceleration, velocity, or displacement time series) of the target structure are recorded over a sufficiently long period, }to ensure stationarity of operating conditions, which is critical for accurate analysis. Ambient forces, including wind, traffic, or thermal effects, can be assumed to act as random broadband inputs, exciting multiple structural modes simultaneously and, thus, allowing to provide a comprehensive representation of the dynamic behaviour. The cross-correlation functions of these response signals are then calculated, either in the time domain or through Cross-Spectral Density in the frequency domain.\cite{Bendat2013} These cross-correlations mimic free vibration response, enabling the identification of modal parameters via standard methods, as though an impulse force had been applied. The assumption of stationary excitations ensures that statistical properties, such as mean and variance, remain constant, allowing the cross-correlation functions to accurately reflect a system dynamic response.

A distinguishing feature of NExT is its reliance on cross-correlation functions between output signals. For two signals $x(t)$ and $y(t)$ measured at different locations, the cross-correlation function is defined as:
\begin{equation}
    R_{xy}(\tau) = \lim_{T \to \infty} \frac{1}{T} \int_0^{T} x(t)y(t+\tau)\, \mathrm{d}t
    \label{eq:next}
\end{equation}
\noindent$R_{xy}(\tau)$ represents the cross-correlation function, which quantifies the similarity between the signals $x(t)$ and $y(t)$ as a function of the time lag $\tau$. Here, $x(t)$ is the time-dependent output signal measured at one location on the structure, and $y(t)$ is the corresponding signal measured at a different location. The parameter $\tau$ denotes the time lag (or shift) between the two signals. The integration is performed over a time window of duration $T$, where $T$ approaches infinity to ensure statistical accuracy. This function replicates the system IRF, enabling modal identification without requiring direct input measurements.

Thus, the idea in this work is to implement NExT on output-only vibration time series, obtain the system approximated IRF, convert it into the frequency domain (using the Fast Fourier Transform, implemented in MATLAB \verb|fft| function\footnote{\url{https://uk.mathworks.com/help/matlab/ref/fft.html}.}), and, finally feed the FRF, {i.e.} the frequency domain counterpart of the IRF, to the LF. On the other hand, only the IRF is needed for ERA, which is not explicitly discussed in this work as it is used solely as a benchmark method. The interested reader is referred to \cite{Pappa1998} for a detailed overview of the method. The NExT and ERA implementations used in this work are retrieved from.\cite{AlRumaithi2024}

\section{Numerical Case Study}\label{sec:num}
\begin{figure}[hbt!]
\centering
		\includegraphics[width=.45\textwidth]{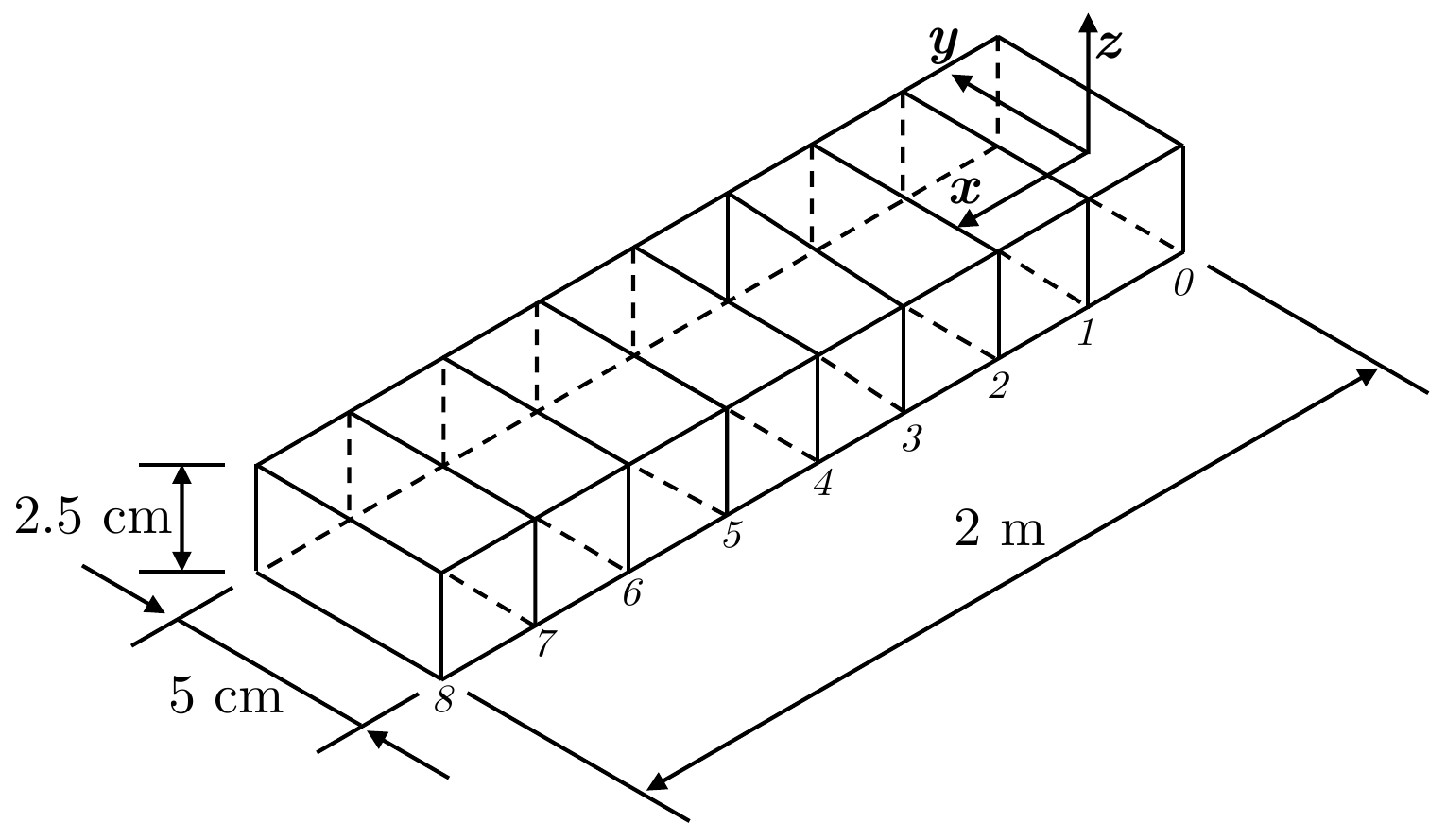}			
	\caption{Numerical case study: 3D Euler Bernoulli beam numerical system.}\label{fig:beam}
\end{figure}

In order to validate the proposed output-only modal identification method, a numerical system of an Euler-Bernuilli cantilever beam is introduced in \cref{fig:beam}. The beam is divided into 8 elements, with the constrained end located at node 0. The beam is made out of aluminium with a density ($\rho$) of 2700  kgm\textsuperscript{-3}, Young’s modulus (E) of 70 GPa and the cross-sectional parameters resulting from the dimensions in \cref{fig:beam}. Modal damping ratio {values of 1 and} 3\% {are} considered for all modes and the mass and stiffness matrices are assembled from standard 2D Euler-Bernoulli beam theory elements (see \cref{eq:mass_matrix,eq:stiffness_matrix} in the Appendix). {This allows us to study the influence of $\zeta_n$ and noise on the identification accuracy by having two numerical models; by having two numerical models; one for relatively low ($\zeta_n$ = 1\%) and the other for relatively high damping ($\zeta_n$ = 3\%).}

\begin{figure}[hbt!]
\centering
		\includegraphics[width=.45\textwidth]{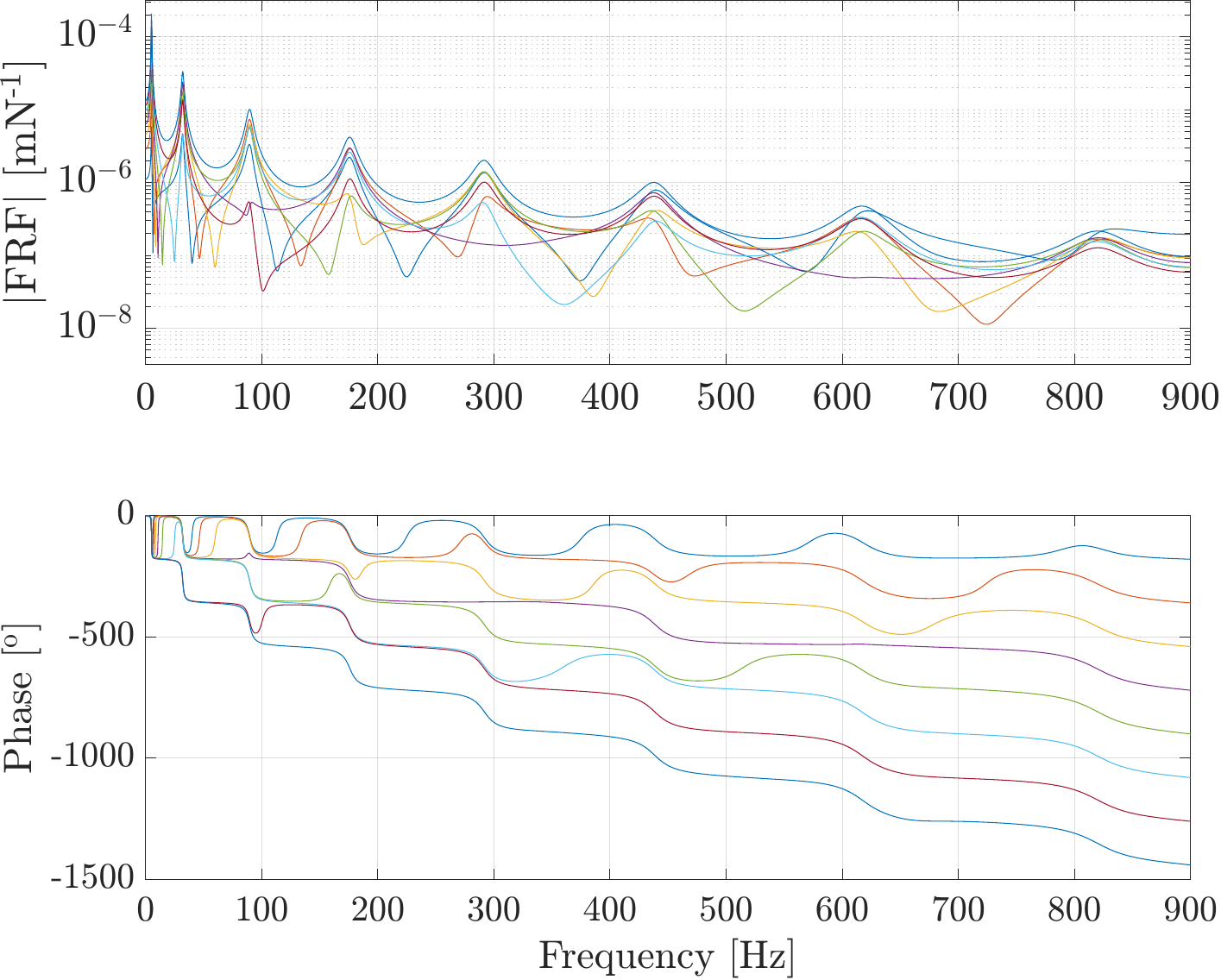}			
	\caption{Numerical case study: FRF of the numerical system.}\label{fig:beamFRF}
\end{figure}

The numerical system is excited with a vertical force of unity, 1 N, at node 1 at the initial time t = 0 s. The response of the system is then recorded for 30 s at a sampling frequency, $f_s$, of 1800 Hz. The selected $f_s$ allows the inspection of all modes below 900 Hz (Nyquist criterion), which in this case are eight. The modes peaks and phase shifts are clearly shown in \cref{fig:beamFRF}.

The baseline analytical modal data are obtained by eigenanalysis of the system mass and stiffness matrices and considering the imposed damping ratio. On the other hand, all the time history responses (8 displacement-only degrees of freedom) of the system are used to compute the system IRF via NExT, using the displacement response measured at node 1 as a reference channel. At that point, the IRF is used as input for ERA (NExT-ERA) and converted to the frequency domain to be used with LF (NExT-LF). On the other hand, the displacement time series are directly fed to SSI. The identification process is carried out over a range of orders ($k=$ [16 100]) and stabilisation diagrams are used to extract the stable{ -- likely physically meaningful -- modes. These are not included here for brevity}. The results, in terms of $\omega_n$, $\zeta_n$, and $\bm{\phi}_n$, {are presented in }\cref{tab:beam1} {for the $\zeta_n$ = 1\% model and in} \cref{tab:beam} {for $\zeta_n$ = 3\%.} The $\bm{\phi}_n$ identification results are presented using the MAC value w.r.t. the {analytical} values. {This dataset does not address closely spaced -- in frequency -- modes as the LF has already shown to outperform established methods in this regard, e.g., least-squares complex exponential on a numerical system} \cite{Dessena2022} {and SSI and Numerical algorithm for (4) subspace state space System IDentification (N4SID) on a full aircraft.}\cite{Dessena2024} 

\begin{table*}[hbt!]
    \caption{{Numerical ($\zeta_n=$ 1\%) case study: Beam analytical and identified $\omega_n$ and $\zeta_n$, and MAC Values between the identified $\bm{\phi}_n$ and the analytical counterpart. In brackets the percentage difference between the identified values and the analytical results.}\label{tab:beam1}}
    \centering
    \resizebox{\textwidth}{!}{  
    \begin{tabular}{lcccccccccc}
\hline
   $\zeta_n=$ 1\% & \multicolumn{4}{c}{\textbf{Natural Frequencies {[}Hz{]}}}                     & \multicolumn{3}{c}{\textbf{Damping Ratio {[}-{]}}}      & \multicolumn{3}{c}{\textbf{MAC (diagonal) Value {[}-{]}}} \\ \cline{2-11} 
\textbf{Mode \#} & \textbf{Analytical} & \textbf{SSI} & \textbf{NExT-ERA} & \textbf{NExT-LF} & \textbf{SSI} & \textbf{NExT-ERA} & \textbf{NExT-LF} & \textbf{SSI}  & \textbf{NExT-ERA}  & \textbf{NExT-LF} \\ \hline
\textbf{1}       & 5.10                & 5.10             & 5.10              & 5.10             & 0.010            & 0.010             & 0.010            & 1                 & 1                  & 1                \\
                 & -                   & (0.00)           & (0.00)            & (0.00)           & (0.00)           & (0.00)            & (0.00)           & -                 & -                  & -                \\
\textbf{2}       & 31.97               & 31.99            & 31.99             & 31.99            & 0.010            & 0.010             & 0.010            & 1                 & 1                  & 1                \\
                 & -                   & (0.01)           & (0.01)            & (0.01)           & (0.00)           & (0.00)            & (0.00)           & -                 & -                  & -                \\
\textbf{3}       & 89.61               & 89.62            & 89.62             & 89.62            & 0.010            & 0.010             & 0.010            & 1                 & 1                  & 1                \\
                 & -                   & (0.01)           & (0.01)            & (0.01)           & (0.00)           & (0.00)            & (0.00)           & -                 & -                  & -                \\
\textbf{4}       & 175.89              & 175.90           & 175.90            & 175.90           & 0.010            & 0.010             & 0.010            & 1                 & 1                  & 1                \\
                 & -                   & (0.01)           & (0.01)            & (0.01)           & (0.00)           & (0.00)            & (0.00)           & -                 & -                  & -                \\
\textbf{5}       & 291.78              & 291.80           & 291.80            & 291.80           & 0.010            & 0.010             & 0.010            & 1                 & 1                  & 1                \\
                 & -                   & (0.01)           & (0.01)            & (0.01)           & (0.00)           & (0.00)            & (0.00)           & -                 & -                  & -                \\
\textbf{6}       & 438.52              & 438.54           & 438.54            & 438.54           & 0.010            & 0.010             & 0.010            & 1                 & 1                  & 1                \\
                 & -                   & (0.01)           & (0.01)            & (0.01)           & (0.00)           & (0.00)            & (0.00)           & -                 & -                  & -                \\
\textbf{7}       & 617.31              & 617.34           & 617.34            & 617.34           & 0.010            & 0.010             & 0.010            & 1                 & 1                  & 1                \\
                 & -                   & (0.01)           & (0.01)            & (0.01)           & (0.00)           & (0.00)            & (0.00)           & -                 & -                  & -                \\
\textbf{8}       & 820.08              & 820.12           & 820.12            & 820.12           & 0.010            & 0.010             & 0.010            & 1                 & 1                  & 1                \\
                 & -                   & (0.01)           & (0.01)            & (0.01)           & (0.00)           & (0.00)            & (0.00)           & -                 & -                  & -                \\ \hline
\end{tabular}}
\end{table*}

\begin{table*}[hbt!]
    \caption{{Numerical  case study: Beam analytical (3\%) and identified $\omega_n$ and $\zeta_n$, and MAC Values between the identified $\bm{\phi}_n$ and the analytical counterpart. In brackets the percentage difference between the identified values and the analytical results.}\label{tab:beam}}
    \centering
    \resizebox{\textwidth}{!}{  
\begin{tabular}{lcccccccccc}
\hline
    $\zeta_n=$ 3\%             & \multicolumn{4}{c}{\textbf{Natural Frequencies {[}Hz{]}}}                                         & \multicolumn{3}{c}{\textbf{Damping Ratio {[}-{]}}}      & \multicolumn{3}{c}{\textbf{MAC (diagonal) Value {[}-{]}}}                   \\ \cmidrule(lr){2-5} \cmidrule(lr){6-8} \cmidrule(lr){9-11}
\textbf{Mode \#} & \textbf{Analytical} & \multicolumn{1}{l}{\textbf{SSI}} & \textbf{NExT-ERA} & \textbf{NExT-LF} & \textbf{SSI} & \textbf{NExT-ERA} & \textbf{NExT-LF} & \multicolumn{1}{l}{\textbf{SSI}} & \textbf{NExT-ERA} & \textbf{NExT-LF} \\ \hline
\textbf{1}       & 5.10                & 5.10                                 & 5.10              & 5.10             & 0.030            & 0.028             & 0.030            & 1                                    & 1                 & 1                \\
                 & -                   & (0.00)                               & (0.00)            & (0.00)           & (0.00)           & (-6.67)           & (0.00)           & -                                    & -                 & -                \\
\textbf{2}       & 31.97               & 31.99                                & 31.99             & 31.99            & 0.030            & 0.030             & 0.030            & 1                                    & 1                 & 1                \\
                 & -                   & (0.06)                               & (0.06)            & (0.06)           & (0.00)           & (0.00)            & (0.00)           & -                                    & -                 & -                \\
\textbf{3}       & 89.57               & 89.62                                & 89.62             & 89.62            & 0.030            & 0.030             & 0.030            & 1                                    & 1                 & 1                \\
                 & -                   & (0.06)                               & (0.06)            & (0.06)           & (0.00)           & (0.00)            & (0.00)           & -                                    & -                 & -                \\
\textbf{4}       & 175.82              & 175.90                               & 175.90            & 175.90           & 0.030            & 0.030             & 0.030            & 1                                    & 1                 & 1                \\
                 & -                   & (0.05)                               & (0.05)            & (0.05)           & (0.00)           & (0.00)            & (0.00)           & -                                    & -                 & -                \\
\textbf{5}       & 291.67              & 291.80                               & 291.80            & 291.80           & 0.030            & 0.030             & 0.030            & 1                                    & 1                 & 0.99             \\
                 & -                   & (0.04)                               & (0.04)            & (0.04)           & (0.00)           & (0.00)            & (0.00)           & -                                    & -                 & -                \\
\textbf{6}       & 438.34              & 438.54                               & 438.54            & 438.54           & 0.030            & 0.030             & 0.030            & 1                                    & 1                 & 0.99             \\
                 & -                   & (0.05)                               & (0.05)            & (0.05)           & (0.00)           & (0.00)            & (0.00)           & -                                    & -                 & -                \\
\textbf{7}       & 617.06              & 617.34                               & 617.24            & 617.34           & 0.030            & 0.030             & 0.030            & 1                                    & 0.94              & 0.97             \\
                 & -                   & (0.05)                               & (0.03)            & (0.05)           & (0.00)           & (0.00)            & (0.00)           & -                                    & -                 & -                \\
\textbf{8}       & 819.75              & 820.12                               & 820.12            & 820.12           & 0.030            & 0.030             & 0.030            & 1                                    & 0.65              & 0.90             \\
                 & -                   & (0.05)                               & (0.05)            & (0.05)           & (0.00)           & (0.00)            & (0.00)           & -                                    & -                 & -                \\ \hline
\end{tabular}}
\end{table*}

{The values identified via NExT-LF, NExT-ERA, and SSI are widely coherent with those from the expected analytical values. In terms of $\omega_n$, the largest deviation, in percentage, is 0.06\% for all methods for modes \#2 and 3 of the $\zeta_n=$ 3\% model. Concerning $\zeta_n$, both methods show a great agreement with the numerical results (deviation $\sim$0\%), except for one case of NExT-ERA ($\zeta_1$ for $\zeta_n=$ 3\%, where the deviation is quite noticeable at -6.67\%). Finally, the identified $\bm{\phi}_n$ MAC values are all close to 1 for both methods. However, this does not hold for the NExT-ERA identified $\bm{\phi}_8$ in the $\zeta_n=$ 3\% model, which is only 0.67; hence, not showing an appropriate correlation with the numerical counterpart. Thus, it can be said that the NExT-LF identification is more robust than NExT-ERA and shows a similar performance to SSI on unperturbed systems.}

\begin{figure*}[hbt!]
    \centering
    \begin{subfigure}[t]{0.28\textwidth}
        \centering
        \caption{}\label{fig:1ab}
        \includegraphics[width=\textwidth]{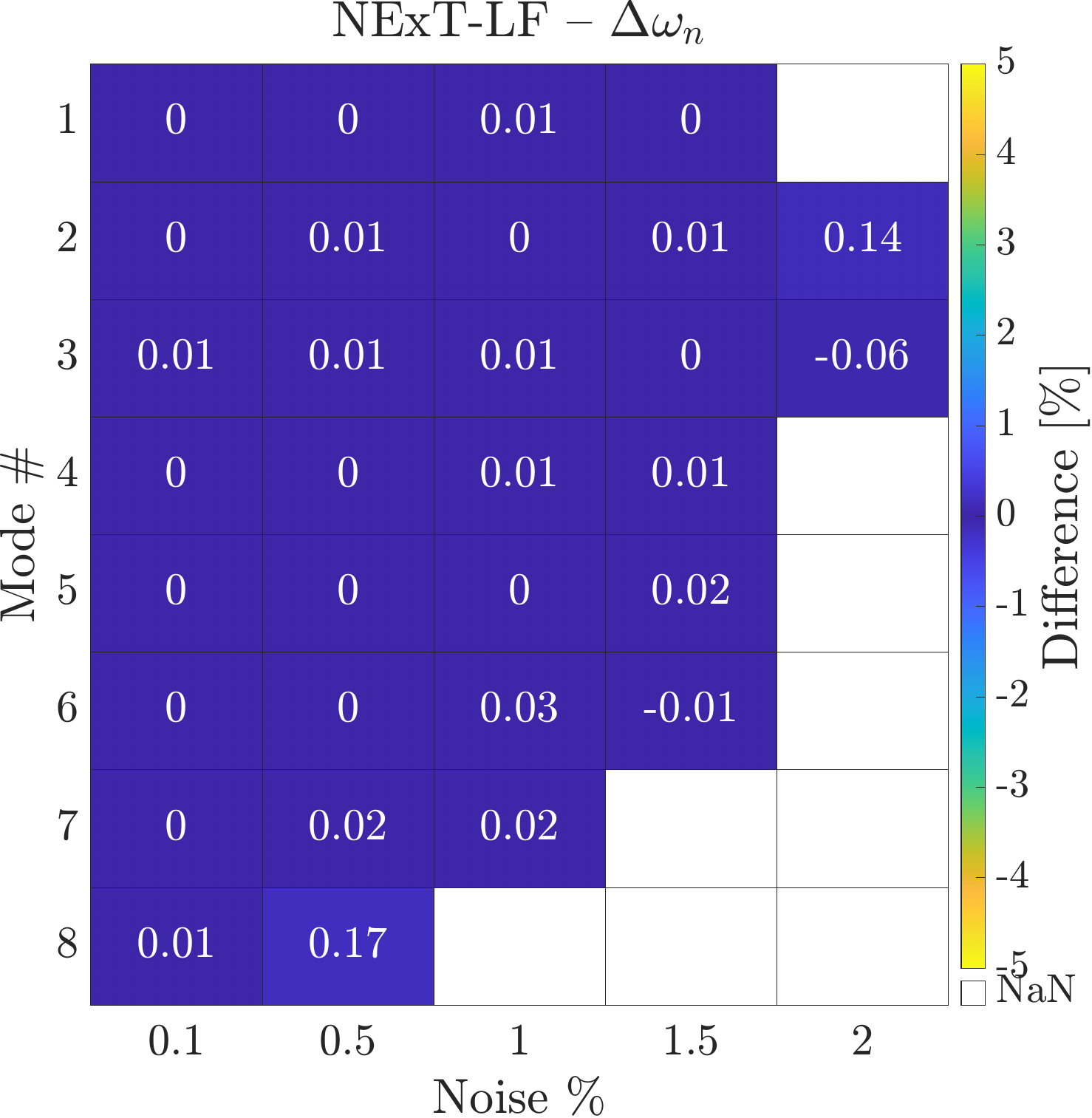}
    \end{subfigure}
    \quad
    \begin{subfigure}[t]{0.28\textwidth}
        \centering
        \caption{}\label{fig:1bb}
        \includegraphics[width=\textwidth]{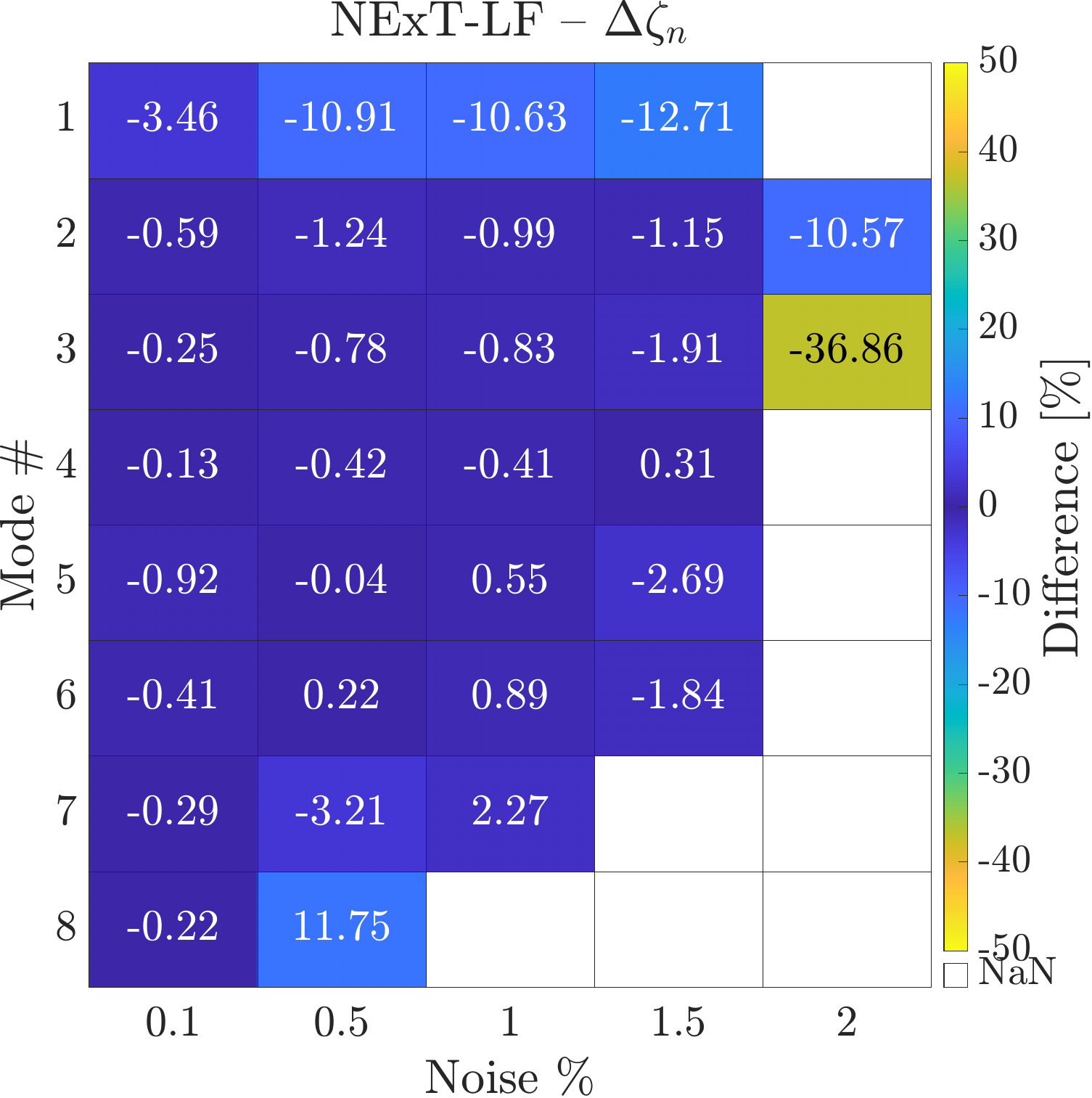}
    \end{subfigure}
    \begin{subfigure}[t]{0.28\textwidth}
        \centering
        \caption{}\label{fig:1cb}
        \includegraphics[width=\textwidth]{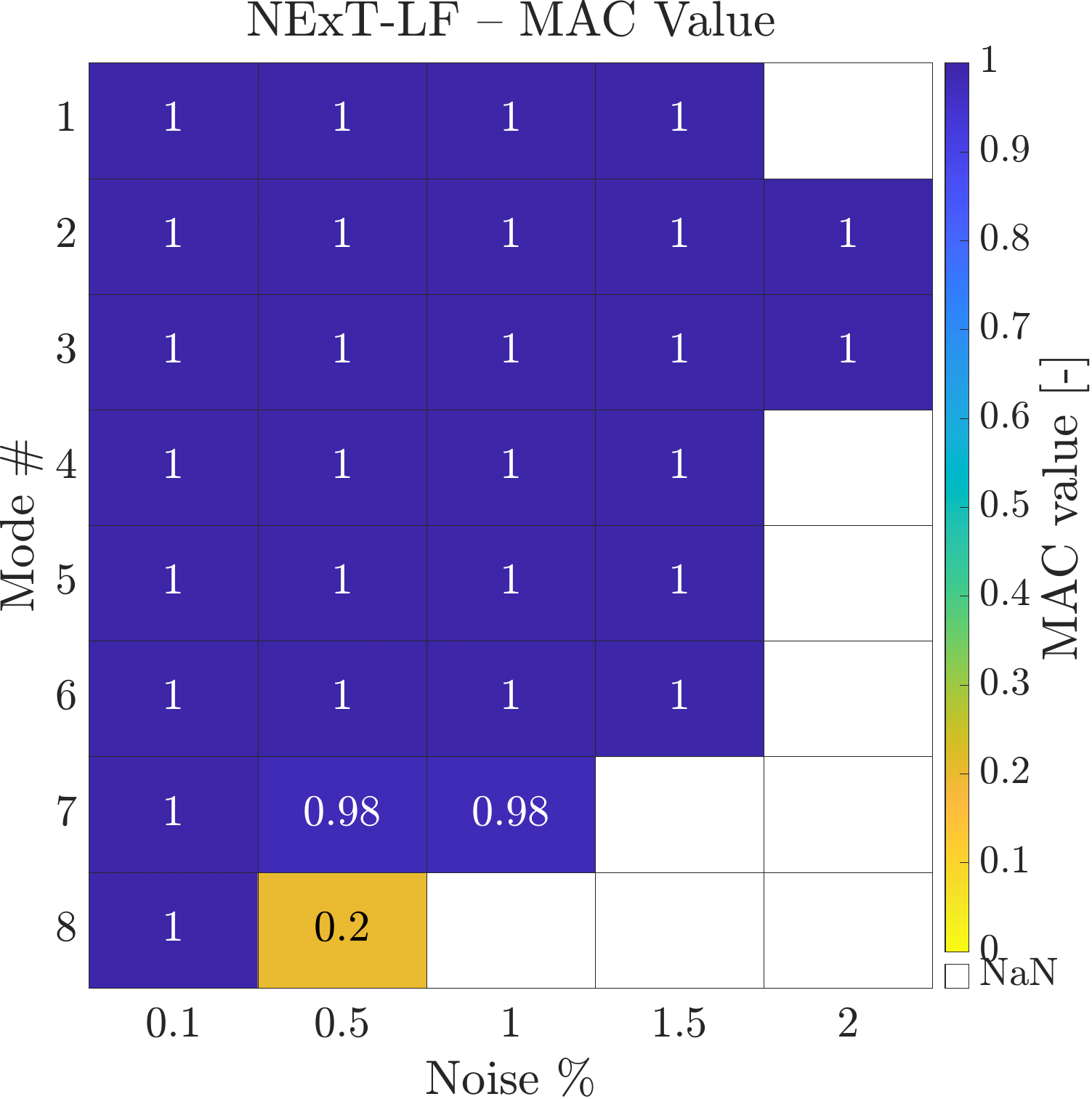}
    \end{subfigure}
    \quad
    \begin{subfigure}[t]{0.28\textwidth}
        \centering
        \caption{}\label{fig:1db}
        \includegraphics[width=\textwidth]{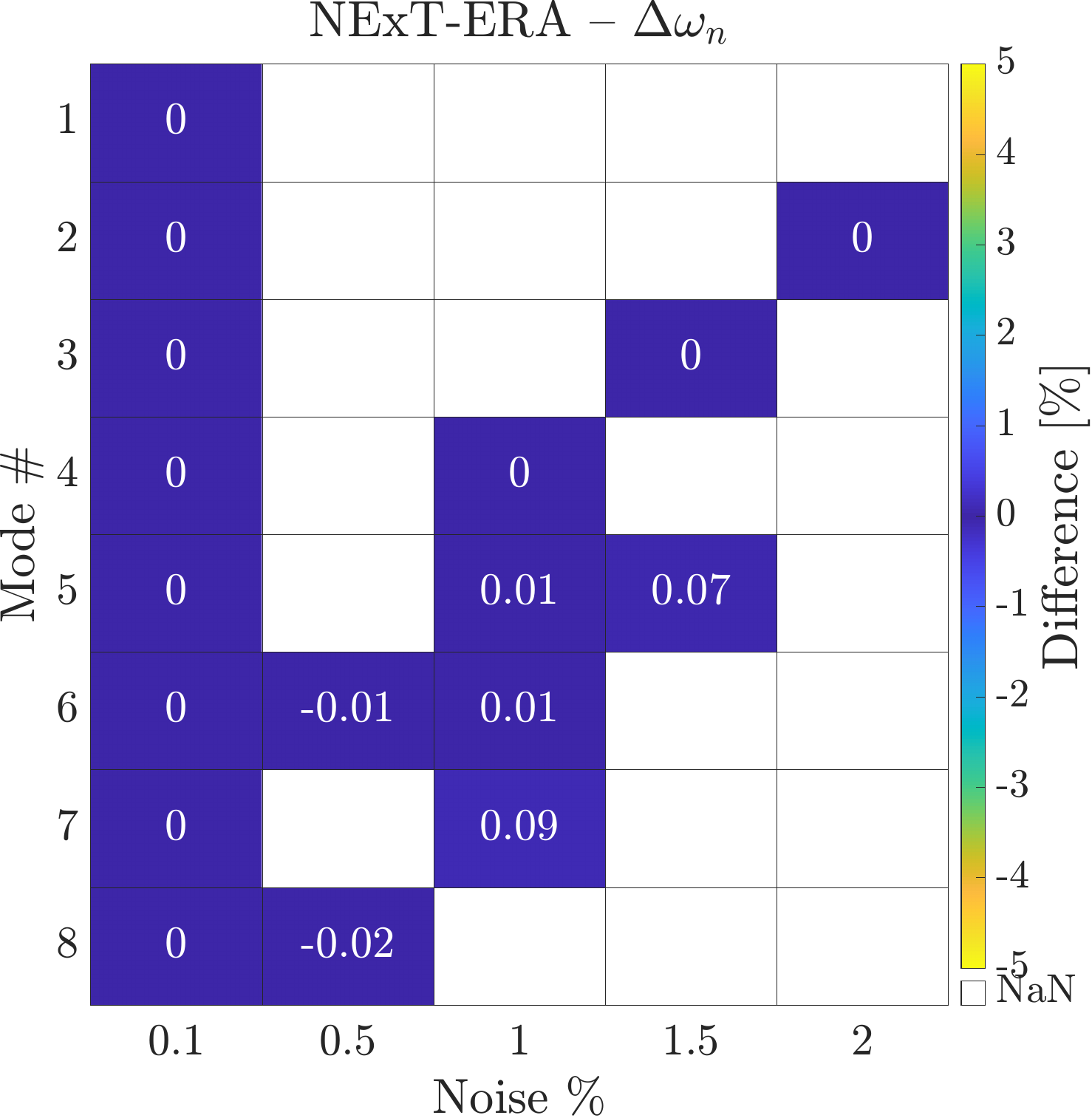}
    \end{subfigure}
    \begin{subfigure}[t]{0.28\textwidth}
        \centering
        \caption{}\label{fig:1eb}
        \includegraphics[width=\textwidth]{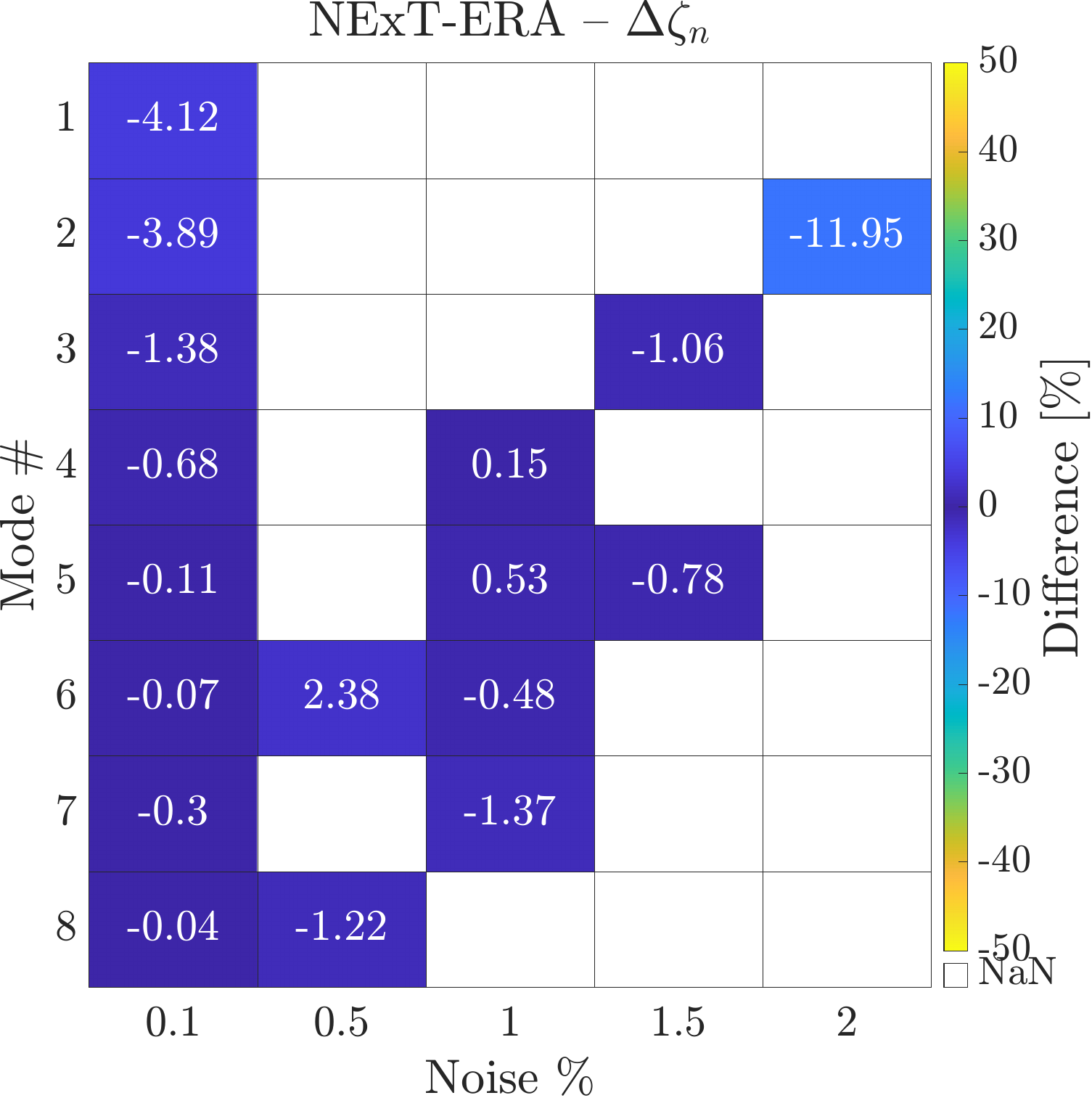}
    \end{subfigure}
    \begin{subfigure}[t]{0.28\textwidth}
        \centering
        \caption{}\label{fig:1fb}
        \includegraphics[width=\textwidth]{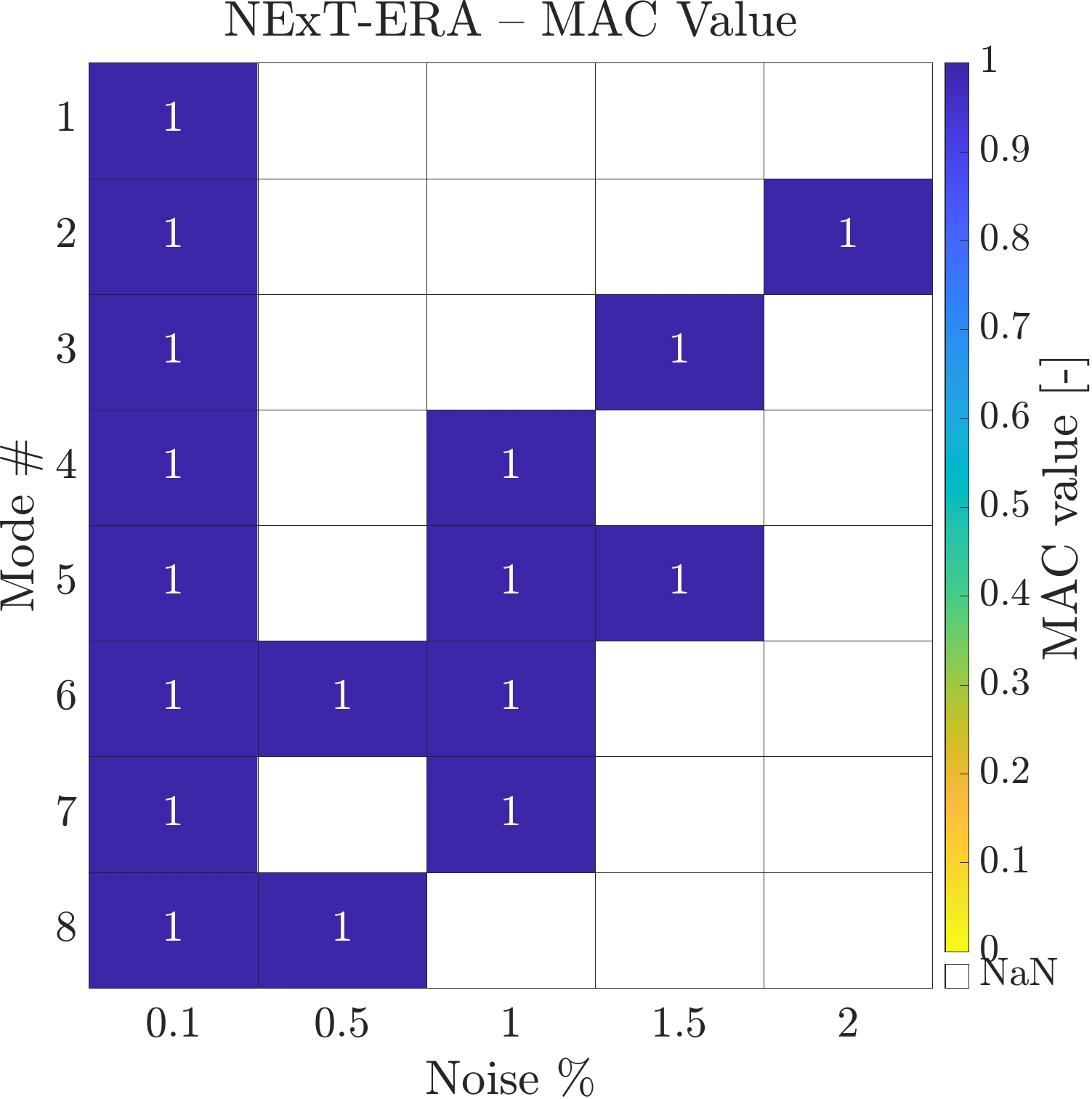}
    \end{subfigure}
        \quad
    \begin{subfigure}[t]{0.28\textwidth}
        \centering
        \caption{}\label{fig:1gb}
        \includegraphics[width=\textwidth]{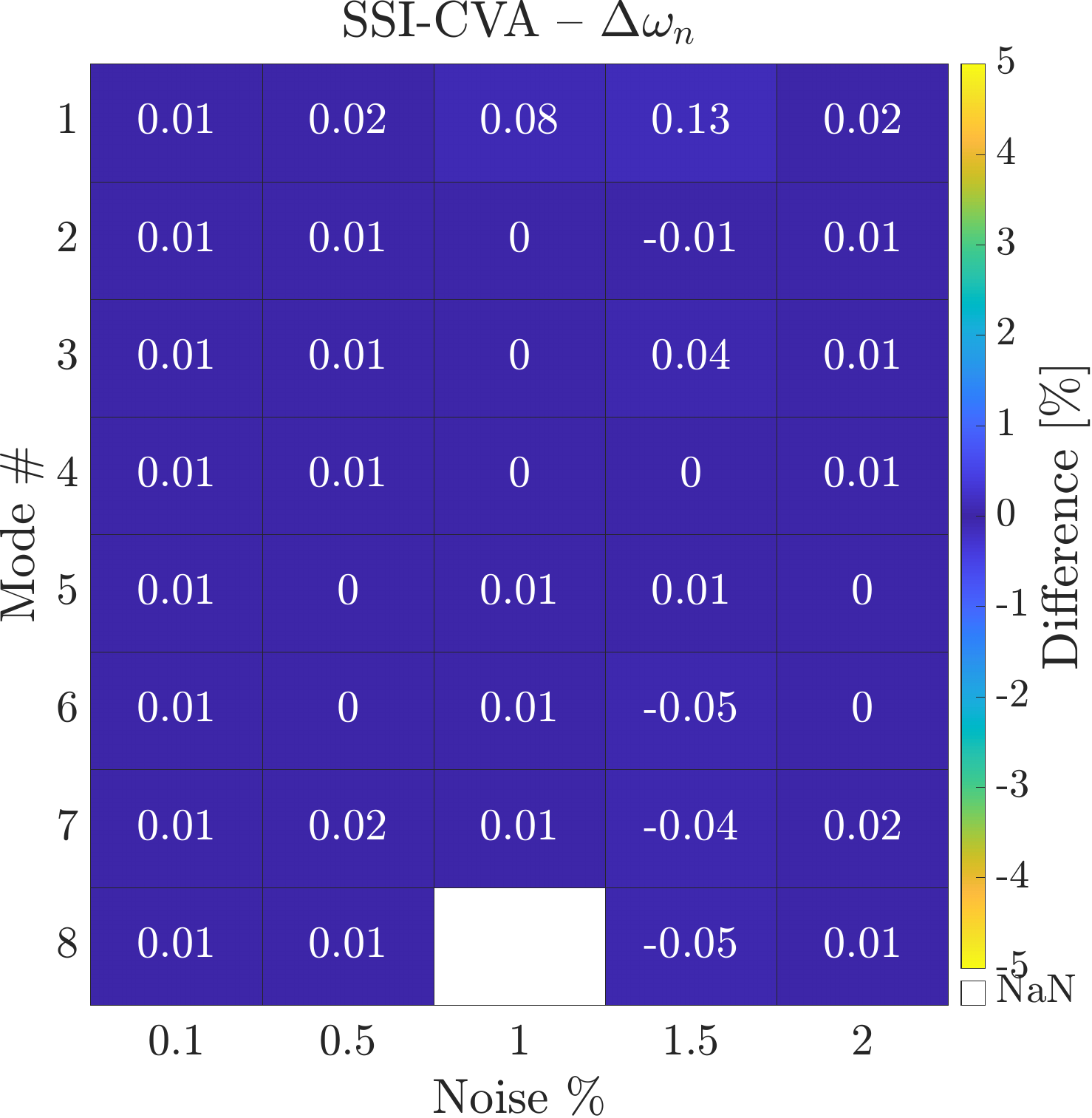}
    \end{subfigure}
    \begin{subfigure}[t]{0.28\textwidth}
        \centering
        \caption{}\label{fig:1hb}
        \includegraphics[width=\textwidth]{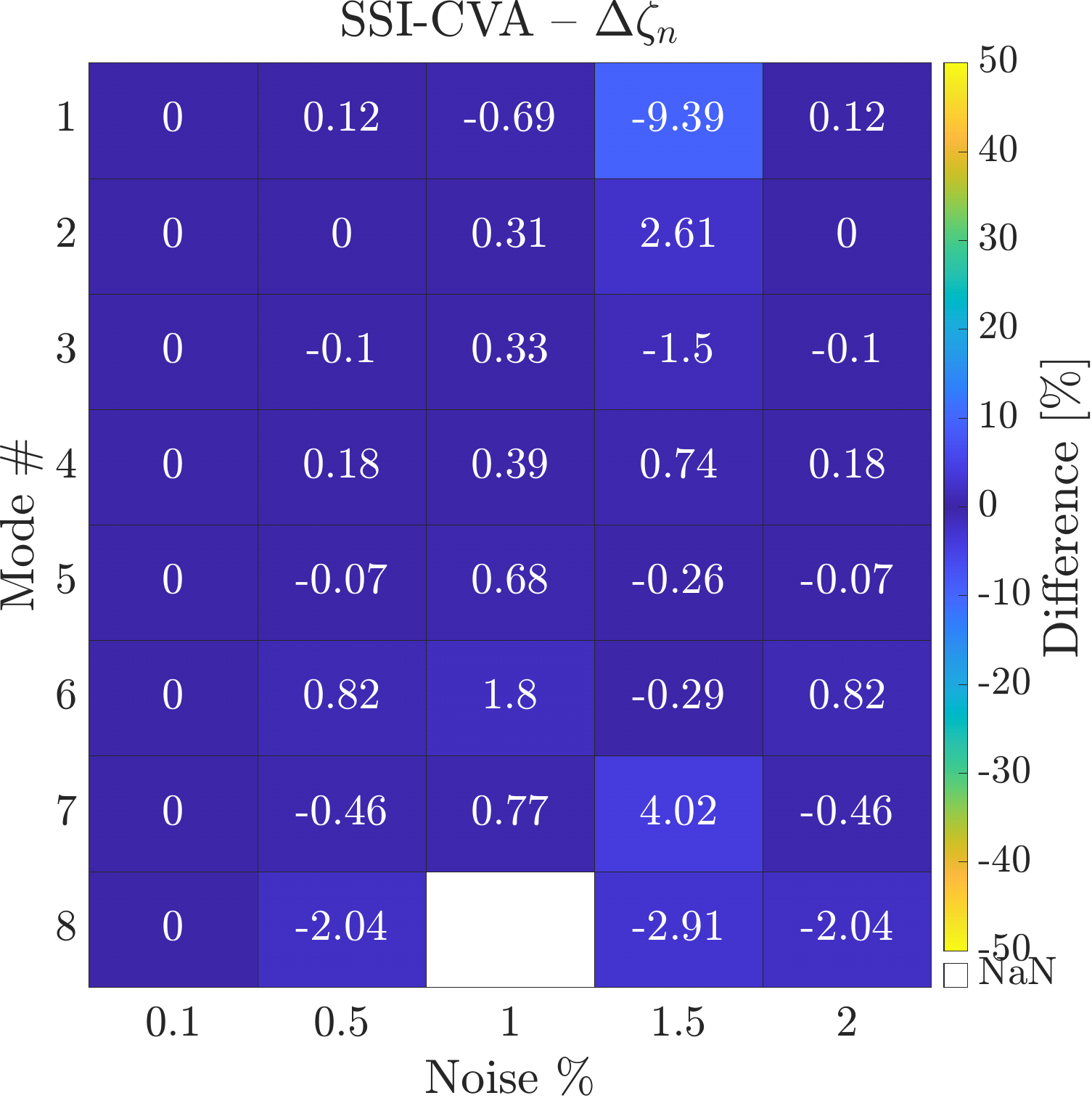}
    \end{subfigure}
    \begin{subfigure}[t]{0.28\textwidth}
        \centering
        \caption{}\label{fig:1ib}
        \includegraphics[width=\textwidth]{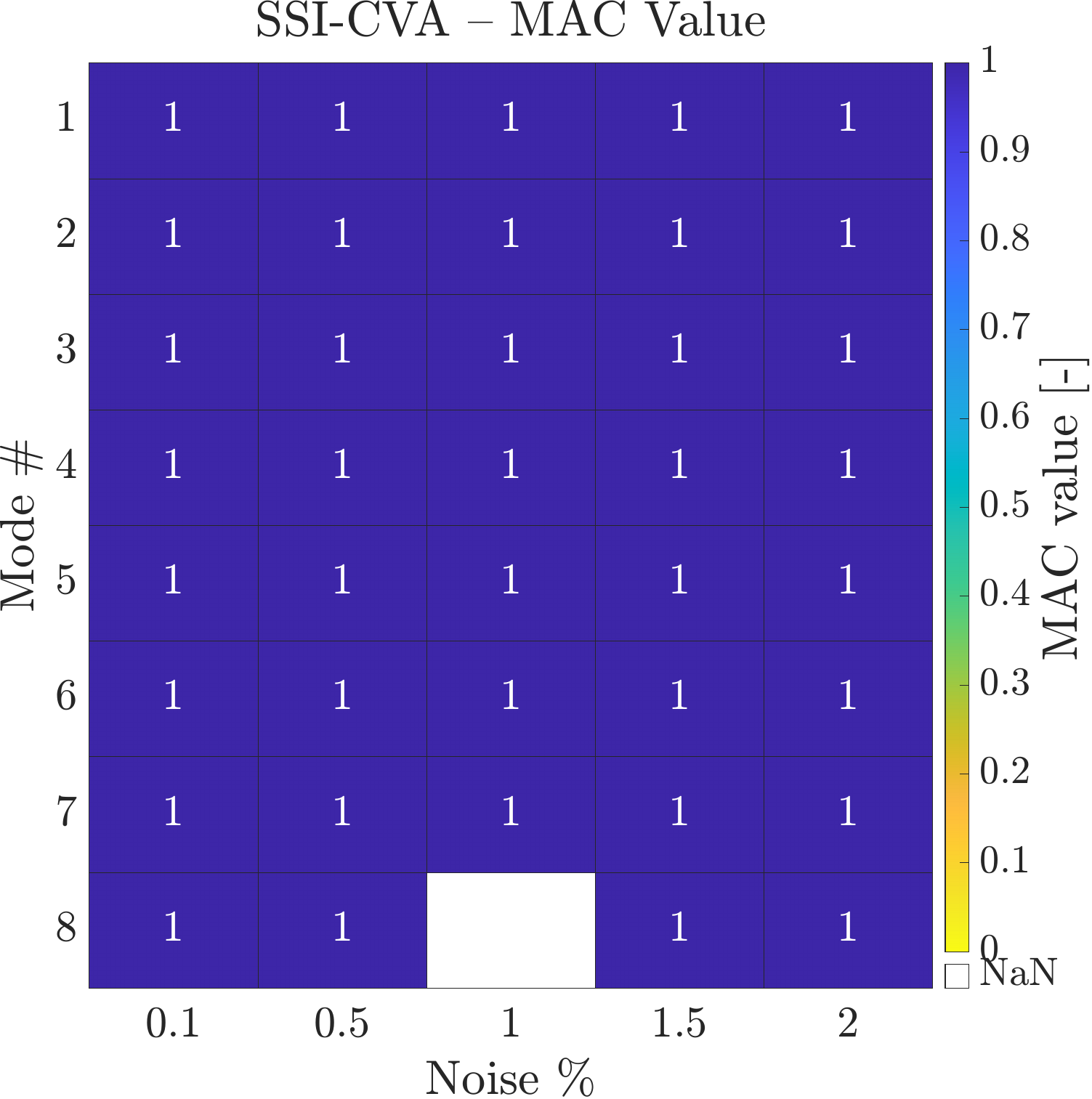}
    \end{subfigure}

    \caption{{Numerical ($\zeta_n=$ 1\%) case study: Effects of input-output noise on the NExT-LF (top row), NExT-ERA (middle row), and SSI (bottom row) identifications (difference w.r.t analytical results) of }$\omega_n$ (\cref{fig:1ab,fig:1db,fig:1gb}), $\zeta_n$ (\cref{fig:1bb,fig:1eb,fig:1hb}), and $\bm{\phi}_n$ -- in terms of MAC Value (\cref{fig:1cb,fig:1fb,fig:1ib}).}\label{fig:num_noise01}
\end{figure*}

\begin{figure*}[hbt!]
    \centering
    \begin{subfigure}[t]{0.28\textwidth}
        \centering
        \caption{}\label{fig:1a}
        \includegraphics[width=\textwidth]{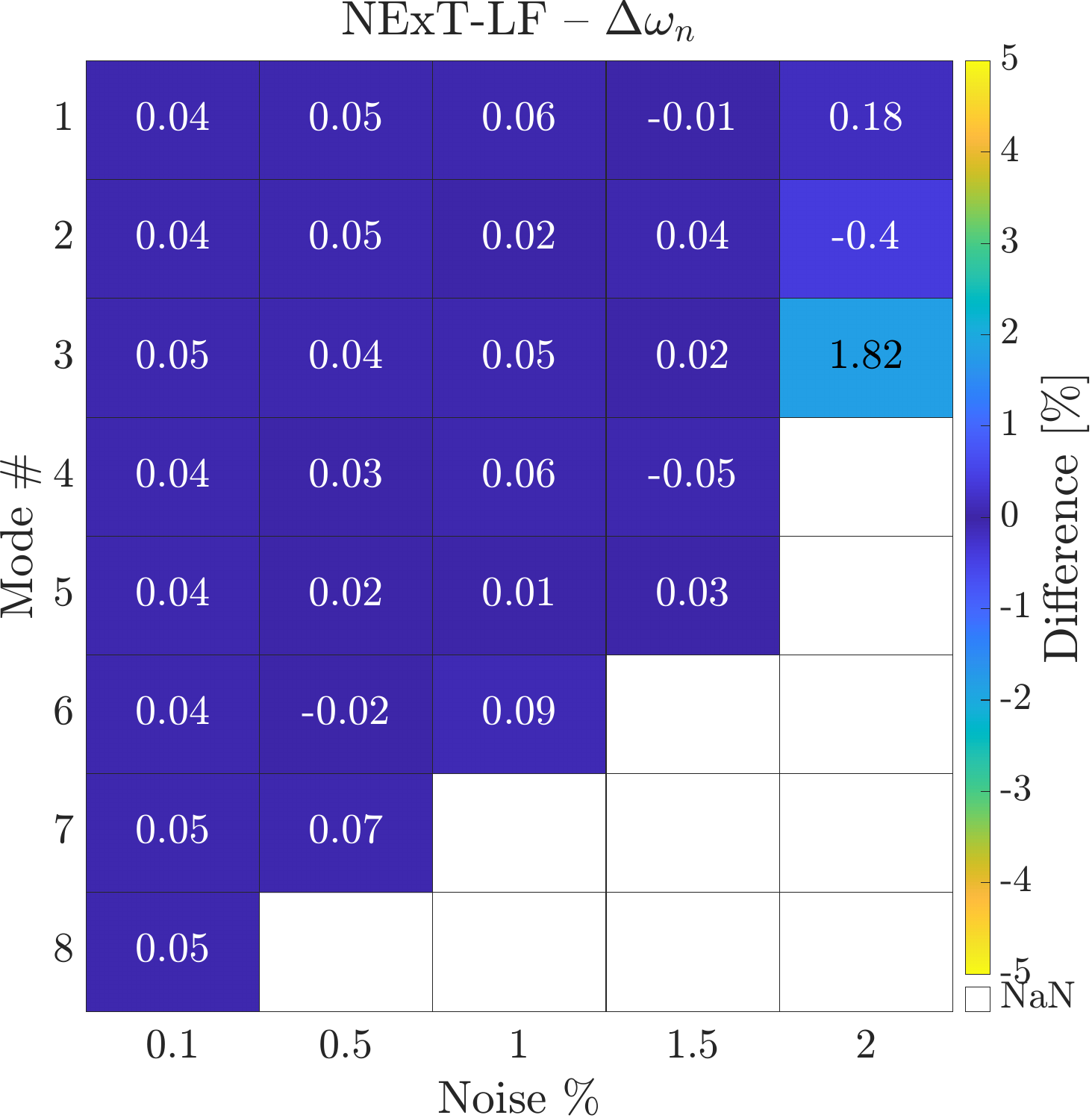}
    \end{subfigure}
    \quad
    \begin{subfigure}[t]{0.28\textwidth}
        \centering
        \caption{}\label{fig:1b}
        \includegraphics[width=\textwidth]{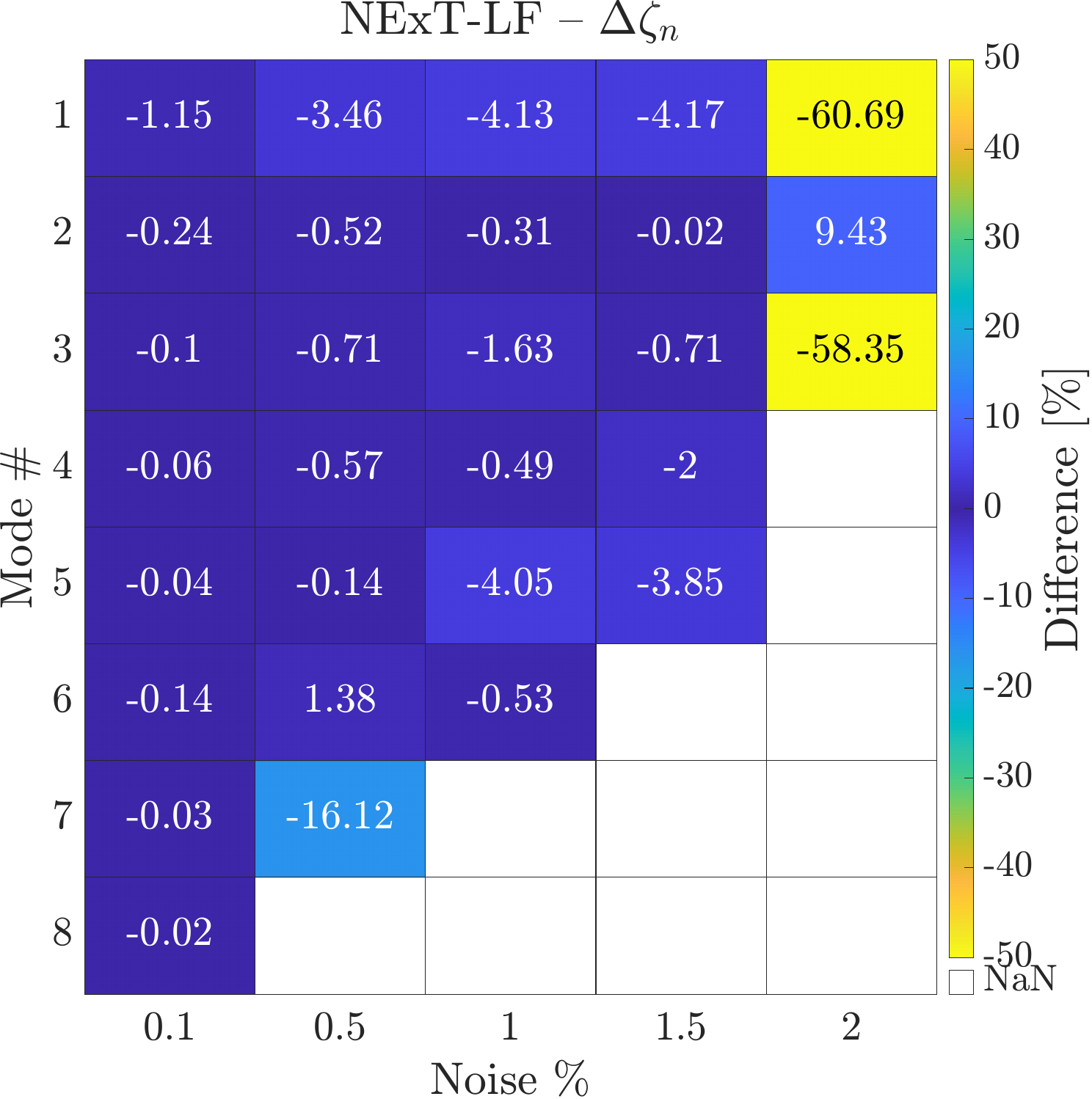}
    \end{subfigure}
    \begin{subfigure}[t]{0.28\textwidth}
        \centering
        \caption{}\label{fig:1c}
        \includegraphics[width=\textwidth]{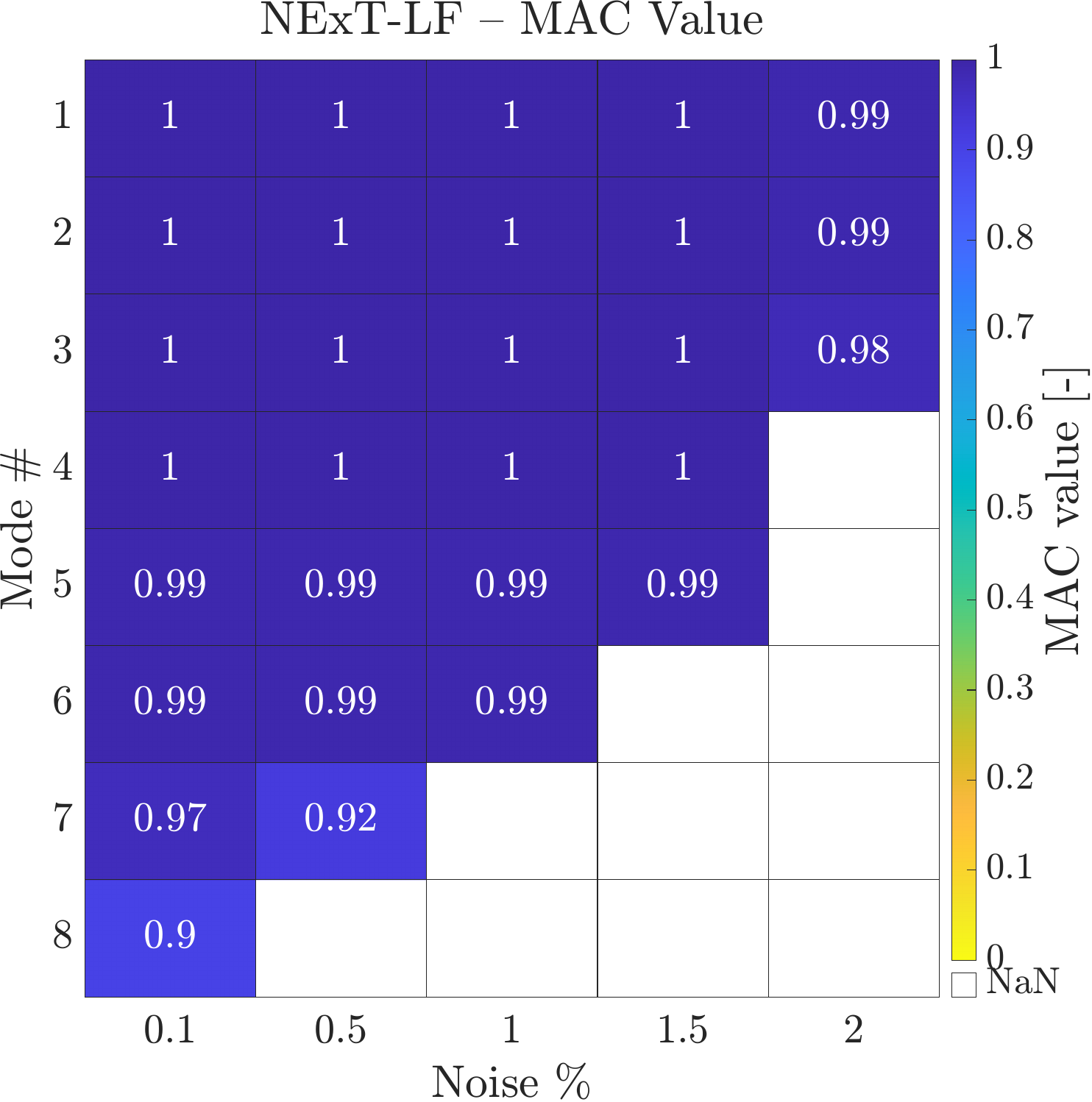}
    \end{subfigure}

    \caption{{Numerical ($\zeta_n=$ 3\%) case study: Effects of input-output noise on the NExT-LF identification (difference w.r.t analytical results) of $\omega_n$ }(\cref{fig:1a}), $\zeta_n$ (\cref{fig:1b}), and $\bm{\phi}_n$ -- in terms of MAC Value (\cref{fig:1c}).}\label{fig:num_noise}
\end{figure*}

However, as per the definition, OMA is often adversely affected by environmental conditions, which in a signal can be usually modelled as noise. Consequently, to assess the newly proposed NExT-LF OMA suitability, the displacement response time series and input force {of the two systems} are corrupted with different levels of additive white Gaussian noise. The level is defined w.r.t the standard deviation of the signal.\footnote{MATLAB notation: \texttt{signal+std(signal.*rand(size(size)*noise\_level}, where \texttt{std} is found at \url{https://uk.mathworks.com/help/matlab/ref/double.std.html} and \texttt{rand} at \url{https://uk.mathworks.com/help/matlab/ref/rand.html}.} In this work, levels of 0.1, 0.5, 1, 1.5, and 2\% are considered. The same identification process described above is carried out for the five resulting noise cases and the results, in terms of deviation from the analytical results, for $\omega_n$ and $\zeta_n$, and MAC values, for $\bm{\phi}_n$, are presented in \cref{fig:num_noise01} and \cref{fig:num_noise} for the $\zeta_n=$ 1\% and $\zeta_n=$ 3\% models respectively. {Please note that only the NExT-LF results are presented in }\cref{fig:num_noise}{ for the $\zeta_n=$ 3\% model, as results similar to those for $\zeta_n=$ 1\% model} (\cref{fig:num_noise}) were found for NExT-ERA and SSI.
\renewcommand{\thesubfigure}{\Alph{subfigure}}

\renewcommand{\thesubfigure}{\Alph{subfigure}}

\Cref{fig:1ab,fig:1db,fig:1gb} {show the difference in percentage ($\Delta$) between the analytical and identified, respectively via NExT-LF, NExT-ERA, and SSI, $\omega_n$ of the $\zeta_n =$ 0.1\% model. It is clear that, while all methods can deliver a full identification for the lowest noise case, the NExT-LF identification is much more robust to noise than that from NExT-ERA, both in terms of the number of modes identified and their accuracy. The same is shown in }\cref{fig:1bb,fig:1eb,fig:1hb} {for $\zeta_n$, but the NExT-LF identified modes with a NExT-ERA counterpart tend to show a slightly higher error, such as for modes \# 3 and 4 for the 0.5\% noise case, which, however, is still well below 1\%. Finally, the MAC values in }\Cref{fig:1cb,fig:1fb,fig:1hb}, {respectively from the NExT-LF, NExt-ERA, and SSI identified $\bm{\phi}_n$, show that all modes identified via NExT-LF are well correlated with the numerical values, showing a minimum MAC value of 0.9. The same cannot be said for NExT-ERA, as the MAC value of $\bm{\phi}_8$ for the 0.1\% case shows a value of 0.65. Nevertheless, SSI can identify all the modal parameters accurately at all noise level, exception made for mode \#8 at 1\% noise, but NExT-LF still performs much better than NExT-ERA. Furthermore, a similar situation is found for the $\zeta_n =$ 0.3\% model, showing that damping does not affect the identification quality in noisy scenarios. This is supported by }\cref{fig:num_noise}, which is very similar to \cref{fig:1ab,fig:1bb,fig:1cb}. {The only difference that the identification of mode \#8 at 0.5\% noise is lost, but mode \#1 at 2\% noise is gained.}

Concluding on the numerical system, it can be asserted that NExT-LF performance, both in terms of accuracy and robustness to artificially-added measurement noise is better than NExT-ERA. This is validated further on an experimental system in the following section.

\section{Operational Modal Analysis of the Sheraton Universal Hotel}\label{sec:exp}
After the successful validation of the NExT-LF method on the numerical dataset, a real-life and -size experimental case study is sought. This is found in the Sheraton Universal Hotel dataset, which, despite {some reported works }for low-cycle damage fatigue and base shear estimation,\cite{Goel2010,Goel2010a,Mantawy2015} does not, to this date and to the best of the authors' knowledge, include any results related to modal parameter identification in the current literature. Thus, this work aims to identify them with the newly proposed NExT-LF and the benchmark method, NExT-ERA.

\begin{figure}[hbt!]
\centering
		\includegraphics[width=.45\textwidth]{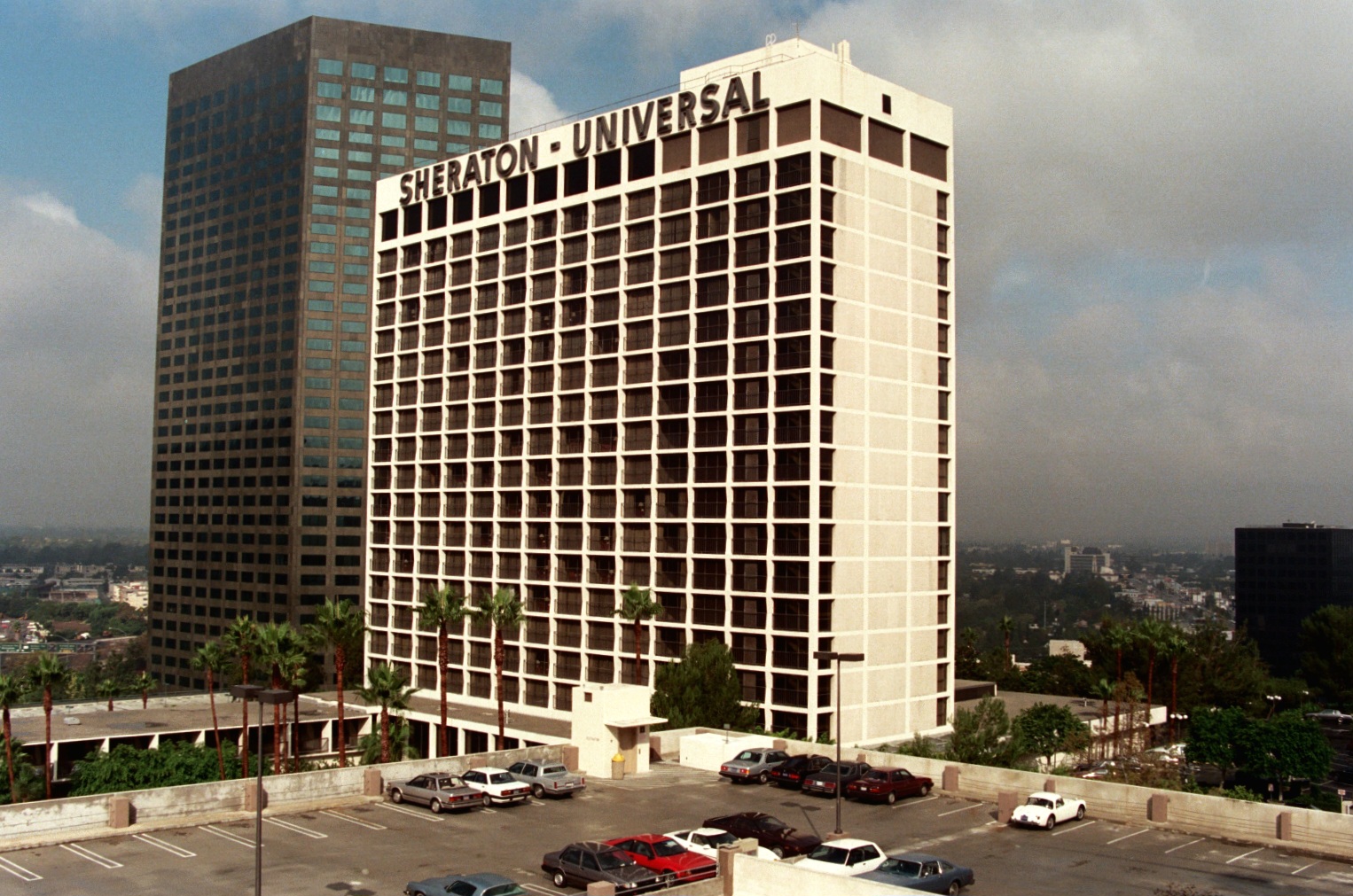}			
	\caption{Experimental case study: Sheraton Universal Hotel -- Universal City, North Hollywood, California (USA) -- in 1987. Location coordinates: 34°08'12.0"N 118°21'36.3"W. Reproduced (Adapted) under terms of the CC-BY license.$^{\text{{REF No. Not Available}}}$ Copyright 2016, The Authors, published by \href{https://www.flickr.com/photos/alan-light/25135256664/in/photostream/}{Flickr, Inc}}\label{fig:hotel}
\end{figure}

The building, constructed in 1967 and shown in \cref{fig:hotel}, stands 173 {ft and 3 in} (54 m) tall. It has a rectangular plan with base dimensions of 96'4" × 198'7" (29.4 × 60.5 m) 
and typical upper floor dimensions of 57'10" × 183'6" (17.6 × 56 m). The vertical load system comprises 4.5–6 in (11.4-15.2 cm) thick concrete slabs supported by reinforced beams and columns, while ductile moment-resisting frames resist lateral forces. Spread footings form the foundation of the structure. The building {has been} permanently instrumented { since the late 90s} as part of the California Strong Motion Instrumentation Program (CSMIP) with accelerometers installed across five floors \cite{cesmd1983}.

Ambient vibration tests were conducted on the 18\textsuperscript{th} of December 1997, by a team from the University of California, Irvine, Stanford University, and Los Alamos National Laboratory \cite{Pardeon1997}. During the ambient vibration tests, excitation was primarily caused by environmental factors such as wind, traffic, and internal mechanical systems. Thirteen Kinemetrics accelerometers were used in conjunction with a Hewlett-Packard 3566A dynamic data acquisition system, enabling the recording and processing of signals. This setup included multiple modules for analog-to-digital conversion and signal processing.

\begin{figure}
\centering
		\includegraphics[width=.45\textwidth]{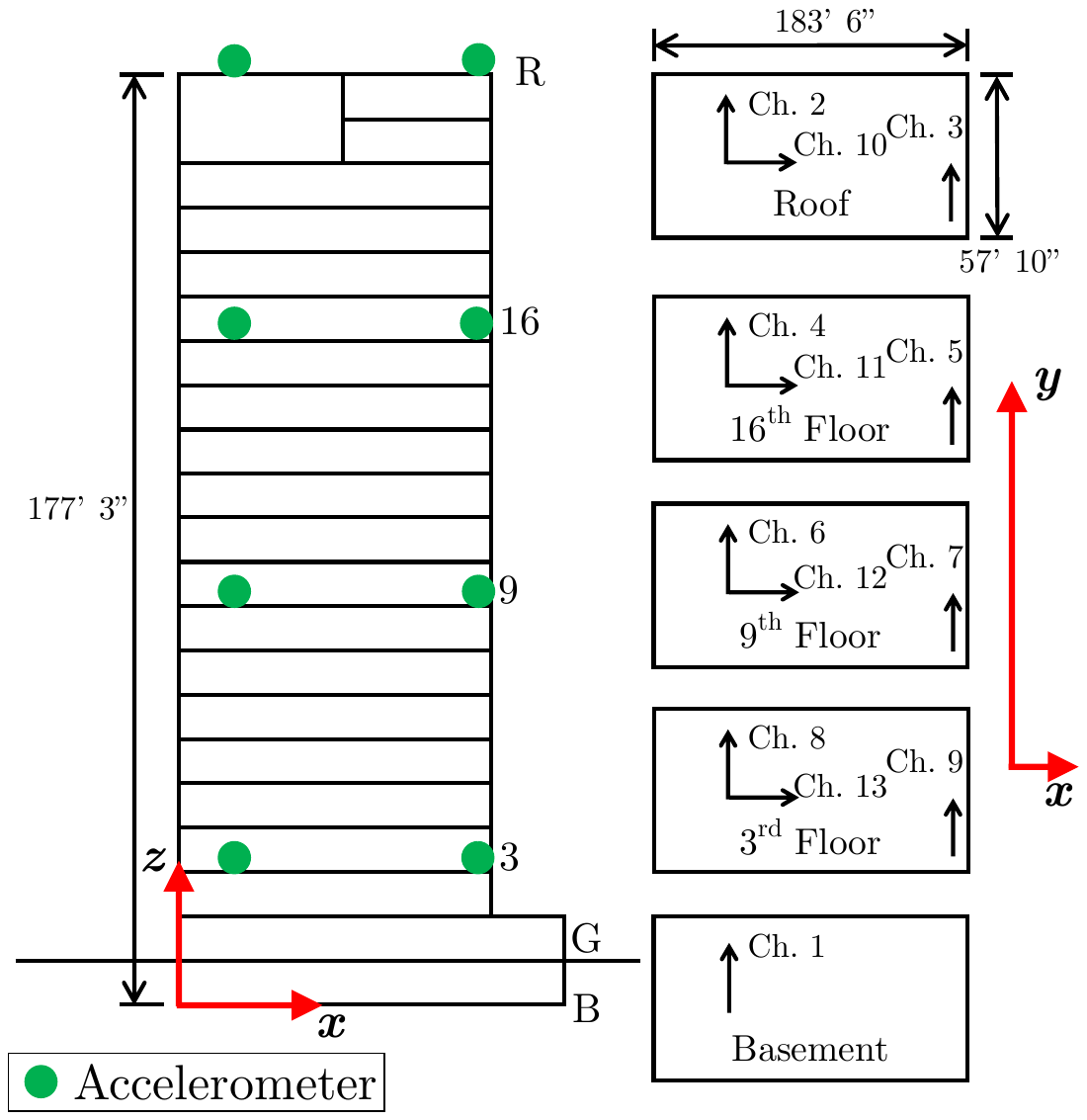}			
	\caption{Experimental case study: Location and direction of the sensors in the building. Green circles outline the accelerometer positions.}
    \label{fig:h_lout}
\end{figure}

The accelerometers were distributed across the building, as shown in \cref{fig:h_lout}, to capture acceleration responses. Sensor positions were based on optimised coverage to assess the structural behaviour effectively. Sampling parameters varied by test designation, with signals categorised into groups based on record length and $f_s$. \Cref{tab:h_data} summarises the test details.
\begin{center}
    \begin{table}[hbt!]%
        \caption{Experimental case study: Sampling parameters and signal length of the ambient vibration tests.\label{tab:h_data}}
        \centering
        \begin{tabular}{lcc}
            \toprule
            \textbf{Test Name} & $\bm{f_s}$ [Hz] & \textbf{Record Length} [s] \\ 
            \midrule
            SH158 & 32 & 64\\
            SH258 & 32 & 64 \\
            SH358 & 16 & 256  \\
            SH458 & 16 & 256 \\
            SH558 & 16 & 512 \\
            SH658 & 16 & 512 \\
            \bottomrule
        \end{tabular}
    \end{table}
\end{center}

The tests also incorporated velocity and displacement measurements from Kinemetrics Ranger Seismometers deployed at specific floors and directions. However, accelerometer-only signals are considered in this work. 
As shown in \cref{fig:h_lout}, basement excluded, there are 12 sensors located on the 3rd, 9th, and 16th floors, as well as on the roof. On each floor, two sensors are oriented North-South (positive direction toward the North), and one sensor perpendicularly (East-West, positive direction towards the East). {With this sensor layout, the system can be identified using the recorded acceleration at this four instrumented floors only, as no additional degrees of freedom are available.}

\begin{figure}[hbt!]
\centering
		\includegraphics[width=.45\textwidth]{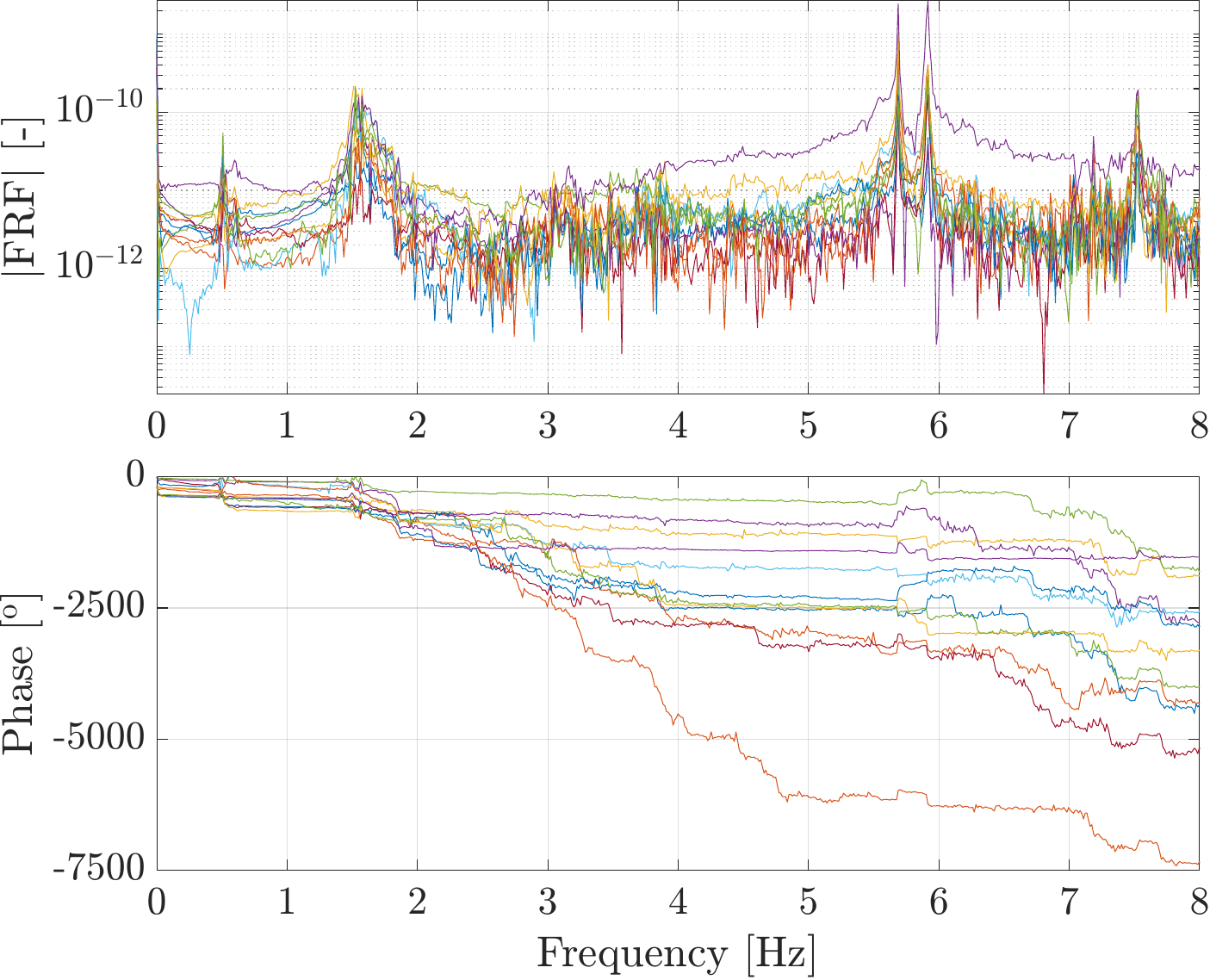}			
	\caption{Experimental case study: Unfiltered FRF of the Sheraton Universal Hotel as computed from the NExT-derived IRF for reference channel 5 and test case SH258.}\label{fig:frfSH}
\end{figure}

Considering a global coordinate system with the origin at the Southwest corner of the building, the $y$-axis points North, the $x$-axis points East, and the $z$-axis represent elevation. The recorded signals are organised in a 12 $\times$ N (N is the mode number) matrix, where:
\begin{itemize}
    \item The first four rows represent acceleration in the $y$ direction,
    \item The next four rows represent acceleration in the $x$ direction,
    \item The final four rows represent rotational signals for each floor, calculated as the difference between two sensors in the $y$-direction multiplied by their distance.
\end{itemize}
For visualisation purposes, the FRF derived from the NExT-derived IRF for reference channel five from test case SH258 is shown in \cref{fig:frfSH}.

\subsection{Results}
Initially, unprocessed signal data were analysed, revealing significant noise interference. To address this, two filters were applied: a band-pass filter for the range of 0.4–9.5 Hz and an empirical Bayesian method with a Cauchy prior. These were implemented in MATLAB using the built-in functions \verb|bandpass|\footnote{\url{https://uk.mathworks.com/help/signal/ref/bandpass.html}.} and \verb|wdenoise|\footnote{\url{https://uk.mathworks.com/help/wavelet/ref/wdenoise.html}.} respectively.

The SI of the Sheraton Hotel was performed using both NExT-LF and NExT-ERA methods. All six datasets (SH158–SH658) were analysed, with different reference channels applied in each analysis to explore all potential outcomes. By evaluating the unfiltered, band-pass filtered, and denoised signals across these datasets, as well as employing five different reference channels per dataset, a total of 90 analyses were conducted. Despite the varying conditions across these analyses, the results for the low-frequency modes were largely consistent. {In all instances, stabilisation diagrams were used to identify the stable modes.}

The NExT-LF and NExT-ERA identification of the signals revealed significant noise, which affected the stability of mode identification. Notably, NExT-ERA showed a recurring tendency to detect spurious modes, while NExT-LF identified fewer stable modes. However, the modes identified by NExT-LF consistently aligned with FRF peaks demonstrate superior robustness to noise. Thus, {the results on this experimental dataset are in agreement with those of the numerical system described} in \cref{sec:num}. This distinction highlights the effectiveness of NExT-LF in minimising the detection of fictitious modes.

Using different reference channels {resulted in} slight variations in the identified modes. Reference channels aligned with the Y direction (e.g., reference channel \#5) tended to result in the identification of more modes. Low-frequency modes were consistently detected within a narrow frequency range across analyses, with examples such as the first mode appearing at 0.64 Hz in one analysis and 0.68 Hz in another, both exhibiting similar mode shapes and damping ratios.

Across all analyses, five stable modes were identified. These modes showed consistency in their natural frequencies ($\omega_n$), damping ratios ($\zeta_n$), and mode shapes ($\bm{\phi}_n$), and both NExT-LF and NExT-ERA methods detected them. These, featuring the full $\bm{\phi}_n$ results following this breakdown described above, are shown in \cref{tab:modal_params_full} in the Appendix, while \cref{tab:sheraton} here shows the MAC value between the NExT-ERA and NExT-LF $\bm{\phi}_n$.

\begin{table}[hbt!]
    \caption{Natural frequencies, damping ratios, and MAC values for modes identified using NExT-ERA and NExT-LF.}
    \centering
    \begin{tabular}{@{}lcccccc@{}}
        \toprule
         & \multicolumn{2}{c}{\textbf{Natural Frequencies [Hz]}} & \multicolumn{2}{c}{\textbf{Damping Ratio [-]}} & \multicolumn{1}{c}{\textbf{MAC Value [-]}} \\
        \cmidrule(lr){2-3} \cmidrule(lr){4-5} \cmidrule(lr){6-6}
        \textbf{Mode \#} & \textbf{NExT-ERA} & \textbf{NExT-LF} & \textbf{NExT-ERA} & \textbf{NExT-LF} & \textbf{NExT-ERA} vs \textbf{NExT-LF}\\
        \midrule
        \textbf{1} & 0.64 & 0.64 & 0.02 & 0.03 & 1 \\
        \textbf{2} & 1.86 & 1.86 & 0.02 & 0.01 & 1 \\
        \textbf{3} & 3.40 & 3.40 & 0.02 & 0.02 & 0.99 \\
        \textbf{4} & 6.02 & 6.01 & 0.02 & 0.02 & 0.93 \\
        \textbf{5} & 7.00 & 6.99 & 0.01 & 0.01 & 0.99 \\
        \bottomrule
    \end{tabular}
    \label{tab:sheraton}
\end{table}

\begin{figure*}[hbt!]
    \centering
    \begin{subfigure}[t]{0.45\textwidth}
        \centering
        \caption{}\label{fig:SHa}
        \includegraphics[width=\textwidth]{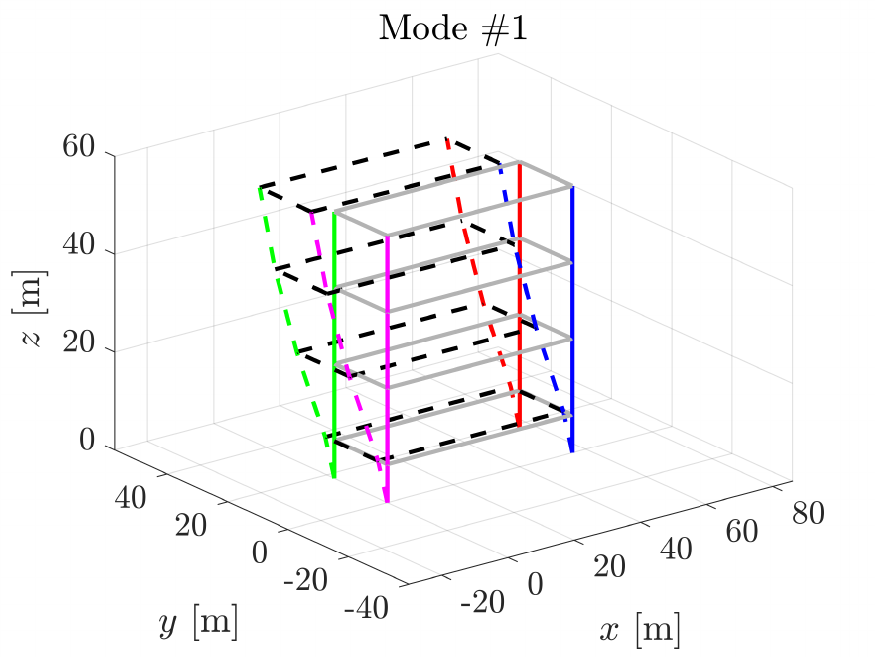}
    \end{subfigure}
    \quad
    \begin{subfigure}[t]{0.45\textwidth}
        \centering
        \caption{}\label{fig:SHb}
        \includegraphics[width=\textwidth]{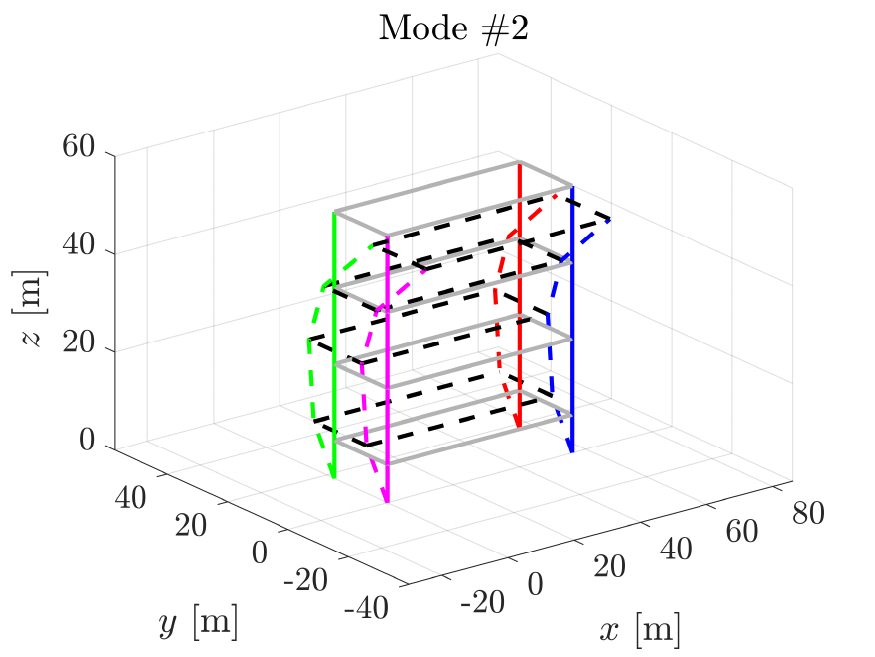}
    \end{subfigure}

    \begin{subfigure}[t]{0.45\textwidth}
        \centering
        \caption{}\label{fig:SHc}
        \includegraphics[width=\textwidth]{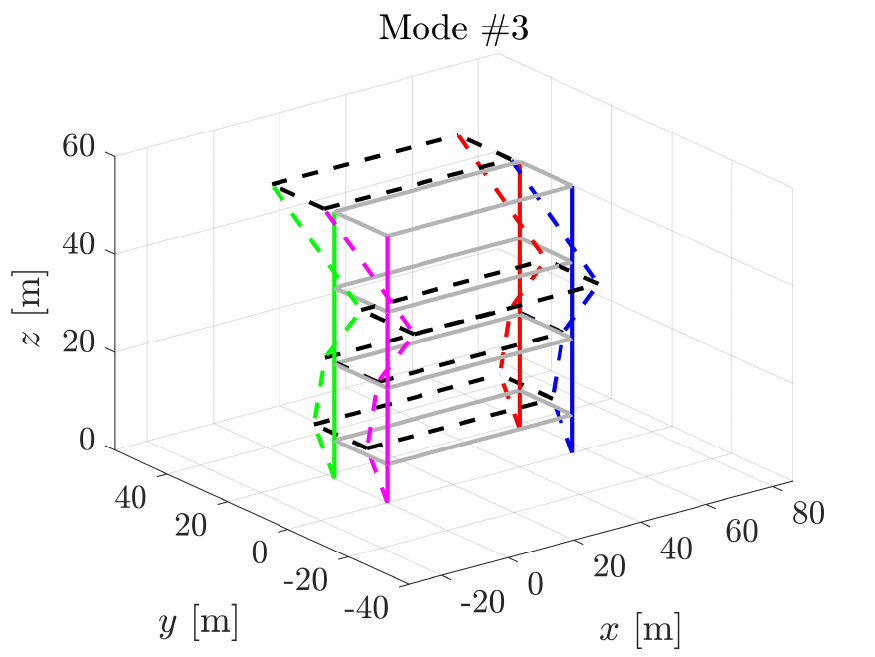}
    \end{subfigure}
    \quad
    \begin{subfigure}[t]{0.45\textwidth}
        \centering
        \caption{}\label{fig:SHd}
        \includegraphics[width=\textwidth]{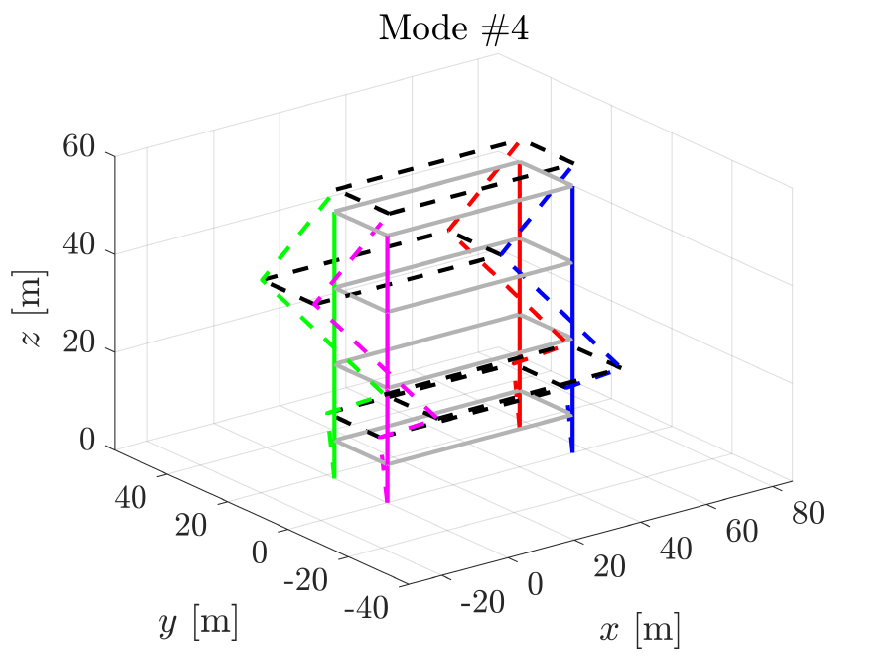}
    \end{subfigure}

    \begin{subfigure}[t]{0.45\textwidth}
        \centering
        \caption{}\label{fig:SHe}
        \includegraphics[width=\textwidth]{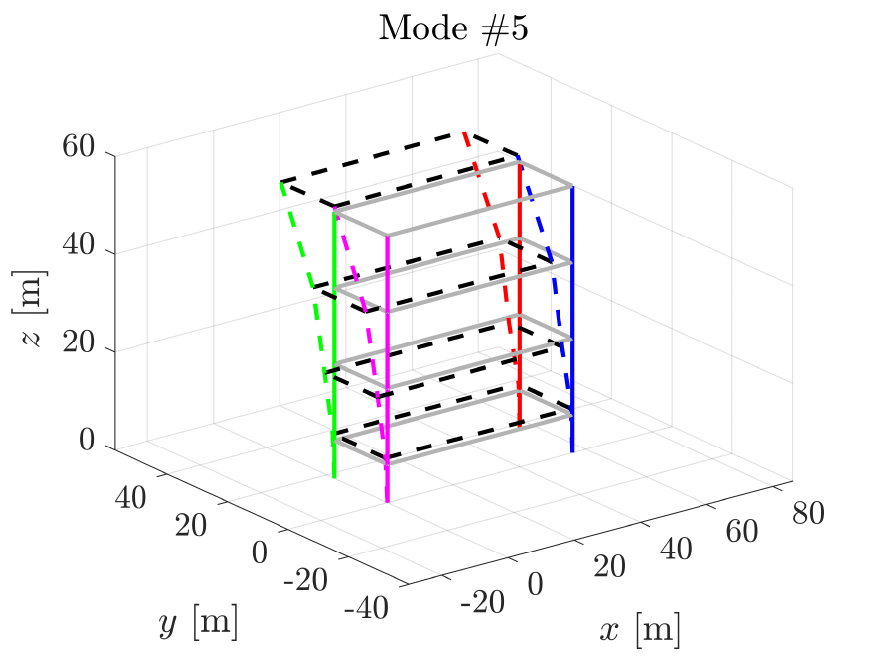}
    \end{subfigure}

    \caption{Experimental case study: $\bm{\phi}_n$ of the Sheraton Universal Hotel first five modes identified via NExT-LF.}\label{fig:SH_modes}
\end{figure*}

Furthermore, the identified $\bm{\phi}_n$ via NExT-LF are shown in \cref{fig:SH_modes}. Only those identified via NExT-LF are presented since the MAC values with the NExT-ERA identified modes are close to 1.  

\renewcommand{\thesubfigure}{\Alph{subfigure}}

The $\bm{\phi}_n$ of the first five modes are consistent with what is expected from a shear-type building such as the one investigated here. Global mode \#1 (0.64 Hz) is a first flexural mode, vibrating in a direction very close to the principal axis for which the moment of inertia is minimum. Probably due to mass and/or stiffness asymmetries in the building, this direction of vibration is not perfectly aligned with the axis along the shorter side of the rectangular plan view and it is slightly skewed. Global modes \#2, \#3, and \#4 (1.86 Hz, 3.49 Hz, and 6.01 Hz, respectively) are the second, third, and fourth flexural modes along the same direction. Finally, the last identified mode, global mode \#5 (6.99 Hz), is the first bending mode in the (stiffer) orthogonal direction, which is, by contrast, occurring almost aligned with the other principal axis of inertia in the horizontal plane. No local modes were identified in the process; again, this is not particularly surprising due to the relatively simple and compact shape of the building, with no appendages, geometric irregularities in vertical or plan view, or soft storeys.

\section{Conclusions}\label{sec:conc}

This work introduced a novel operational modal analysis method, known as NExT-LF, that combines the Natural Excitation Technique (NExT) with the Loewner Framework (LF). The proposed method was validated through numerical and experimental case studies, demonstrating significant improvements in accuracy and robustness to noise compared to traditional methods such as NExT-ERA (NExT with the Eigensystem Realization Algorithm). The following conclusions can be drawn:

\begin{itemize}
    \item NExT-LF identified stable modal parameters under varying noise conditions, outperforming NExT-ERA in terms of reliability, in the noise-robustness sense, and accuracy.
    \item The modal parameters identified from the numerical system showed minimal deviations from analytical values, with NExT-LF achieving a better alignment than NExT-ERA{ and similar performance to tochastic Subspace Identification} to the expected analytical values.
    \item This study presented the first-ever modal parameter identification for the Sheraton Universal Hotel, successfully applying NExT-LF to a real-world operational scenario.
\end{itemize}

The proposed method offers a robust framework for operational modal analysis, with potential applications in structural health monitoring and operational identification of aeronautical systems, e.g. wings in a wind tunnel test. {Further validation will be sought from experimental case studies of output-only data with interesting closely-spaced modes.}


\section*{CRediT authorship contribution statement}
\noindent {Conceptualisation, G.D. and M.C.; methodology, G.D., M.C., and A.Y.; software, G.D. and A.Y.; validation, G.D., M.C., and A.Y.; formal analysis, G.D. and A.Y.; investigation, G.D. and A.Y.; resources,  G.D. and M.C.; data curation, G.D., M.C., and A.Y.; writing---original draft preparation, G.D. and M.C.; writing---review and editing, G.D., M.C., and C.S.; visualisation, G.D., M.C., and A.Y.; supervision, G.D. and M.C., and C.S., funding acquisition, G.D. and M.C..}

\section*{Acknowledgments}
The first author has been supported by the Madrid Government (\emph{Comunidad de Madrid} - Spain) under the Multiannual Agreement with the Universidad Carlos III de Madrid (\href{https://researchportal.uc3m.es/display/act564873}{IA\_aCTRl-CM-UC3M}) and by the Grants for research activity of young PhD holders, part of the UC3M Own Research Program (\href{https://researchportal.uc3m.es/display/act571737}{\emph{Ayudas para la Actividad Investigadora de los Jóvenes Doctores, del Programa Propio de Investigación de la UC3M}}). The second author is supported by the \emph{Centro Nazionale per la Mobilità Sostenibile} (MOST - Sustainable Mobility Center), Spoke 7 (Cooperative Connected and Automated Mobility and Smart Infrastructures), Work Package 4 (Resilience of Networks, Structural Health Monitoring and Asset Management).

The Authors thank the \href{https://www.strongmotioncenter.org/}{Center for Engineering Strong Motion Data (CESMD)} for making the Sheraton Universal City Ambient Vibration Tests dataset publicly available.

\section*{Conflict of interest statement}
{The authors declare no potential conflict of interest.}

\section*{Data Availability Statement}\label{sec:data}
The data that support the findings of this study (Data supporting: NExT-LF: A Novel Operational Modal Analysis Method via Tangential Interpolation -- Data file 1) are openly available in Zenodo at \url{http://doi.org/10.5281/zenodo.14975925}. In addition, this study used third-party experimental data (Sheraton Universal City Ambient Vibration Tests -- Data file 2) that the authors do not have permission to share\footnote{As of the writing of this article (11\textsuperscript{th} of December 2024), the dataset is not available anymore from the CESMD website. The data used in this work was retrieved on the 1\textsuperscript{st} of January 2022.}.
Data file 1 is available under the terms of the [GNU General Public License v3.0 (GPL 3.0)].


\appendix

\section{Beam Element Matrices\label{sec:app1}}
The Euler-Bernoulli beam element mass and stiffness matrices are presented, respectively, in \cref{eq:mass_matrix,eq:stiffness_matrix}.
\begin{equation}
\mathbf{M_e} = \frac{\rho A L}{420}
\begin{bmatrix}
156 & 22L & 54 & -13L \\
22L & 4L^2 & 13L & -3L^2 \\
54 & 13L & 156 & -22L \\
-13L & -3L^2 & -22L & 4L^2
\end{bmatrix}
\label{eq:mass_matrix}
\end{equation}

\begin{equation}
\mathbf{K_e} = \frac{EI}{L^3}
\begin{bmatrix}
12 & 6L & -12 & 6L \\
6L & 4L^2 & -6L & 2L^2 \\
-12 & -6L & 12 & -6L \\
6L & 2L^2 & -6L & 4L^2
\end{bmatrix}
\label{eq:stiffness_matrix}
\end{equation}
where $\rho$ is the material density, $A$ is the cross-sectional area, $L$ is the element length, and $EI$ is the bending stiffness (Young's modulus $\times$ second moment of area).

\section{Sheraton Universal Hotel Raw Identification Results\label{sec:app2}}
The raw identification results of the Sheraton Universal Hotel are shown in \cref{tab:modal_params_full}.
\begin{table*}[hbt!]
    \caption{Modal parameters of Sheraton Universal Hotel identified using NExT-ERA and NExT-LF.}
    \centering
    \resizebox{\textwidth}{!}{  
    \begin{tabular}{@{}lcccccccccc@{}}
        \toprule
        \textbf{Parameter} & \multicolumn{2}{c}{\textbf{Mode \#1}} & \multicolumn{2}{c}{\textbf{Mode \#2}} & \multicolumn{2}{c}{\textbf{Mode \#3}} & \multicolumn{2}{c}{\textbf{Mode \#4}} & \multicolumn{2}{c}{\textbf{Mode \#5}} \\
        \cmidrule(lr){2-3} \cmidrule(lr){4-5} \cmidrule(lr){6-7} \cmidrule(lr){8-9} \cmidrule(lr){10-11}
        & \textbf{NExT-ERA} & \textbf{NExT-LF} & \textbf{NExT-ERA} & \textbf{NExT-LF} & \textbf{NExT-ERA} & \textbf{NExT-LF} & \textbf{NExT-ERA} & \textbf{NExT-LF} & \textbf{NExT-ERA} & \textbf{NExT-LF} \\
        \midrule
        $\omega_n$ [Hz] & 0.64 & 0.64 & 1.86 & 1.86 & 3.40 & 3.40 & 6.02 & 6.01 & 7.00 & 6.99 \\
        $\zeta_n$ [-] & 0.02 & 0.03 & 0.02 & 0.01 & 0.02 & 0.02 & 0.02 & 0.02 & 0.01 & 0.01 \\
        $\bm{\phi}_n$ (y3) & 0 & 0 & 0.02 & 0.01 & 0.01 & 0.03 & 1 & -0.66 & 0.11 & 0.16 \\
        $\bm{\phi}_n$ (y9) & -0.01 & -0.01 & 0.03 & 0.02 & 0.01 & 0.02 & -0.95 & 1 & -0.05 & -0.14 \\
        $\bm{\phi}_n$ (y16) & -0.01 & -0.02 & 0 & 0 & -0.02 & -0.04 & 0.83 & -0.62 & 0.06 & 0.14 \\
        $\bm{\phi}_n$ (yR) & -0.02 & -0.02 & -0.04 & -0.04 & 0.04 & 0.06 & 0.07 & -0.5 & 1 & 1 \\
        $\bm{\phi}_n$ (x3) & 0 & 0 & 0.01 & 0 & 0 & 0 & 0.37 & -0.5 & 0.09 & 0.11 \\
        $\bm{\phi}_n$ (x9) & 0 & 0 & 0.02 & -0.01 & 0 & 0 & -0.05 & 0.16 & -0.26 & -0.27 \\
        $\bm{\phi}_n$ (x16) & 0 & 0 & 0 & 0 & 0.01 & 0 & -0.36 & 0.53 & -0.16 & -0.2 \\
        $\bm{\phi}_n$ (xR) & 0 & 0 & -0.02 & 0.01 & 0 & 0 & 0.43 & -0.48 & 0.06 & 0.11 \\
        $\bm{\phi}_n$ (r3) & 0.07 & 0.13 & -0.35 & -0.37 & -0.7 & -0.74 & -0.02 & 0.01 & 0 & -0.01 \\
        $\bm{\phi}_n$ (r9) & 0.42 & 0.42 & -0.89 & -0.94 & -0.11 & -0.31 & 0.02 & -0.02 & 0 & 0.01 \\
        $\bm{\phi}_n$ (r16) & 0.75 & 0.66 & -0.13 & -0.24 & 1 & 1 & -0.02 & 0.01 & 0.01 & -0.01 \\
        $\bm{\phi}_n$ (rR) & 1 & 1 & -0.89 & -0.77 & 0 & 0 & 0 & 0 & 0.02 & 0.03 \\
        \bottomrule
    \end{tabular}}
    \label{tab:modal_params_full}
\end{table*}

\clearpage

\section*{Author Biography}

\begin{biography}{\includegraphics[width=66pt,height=86pt]{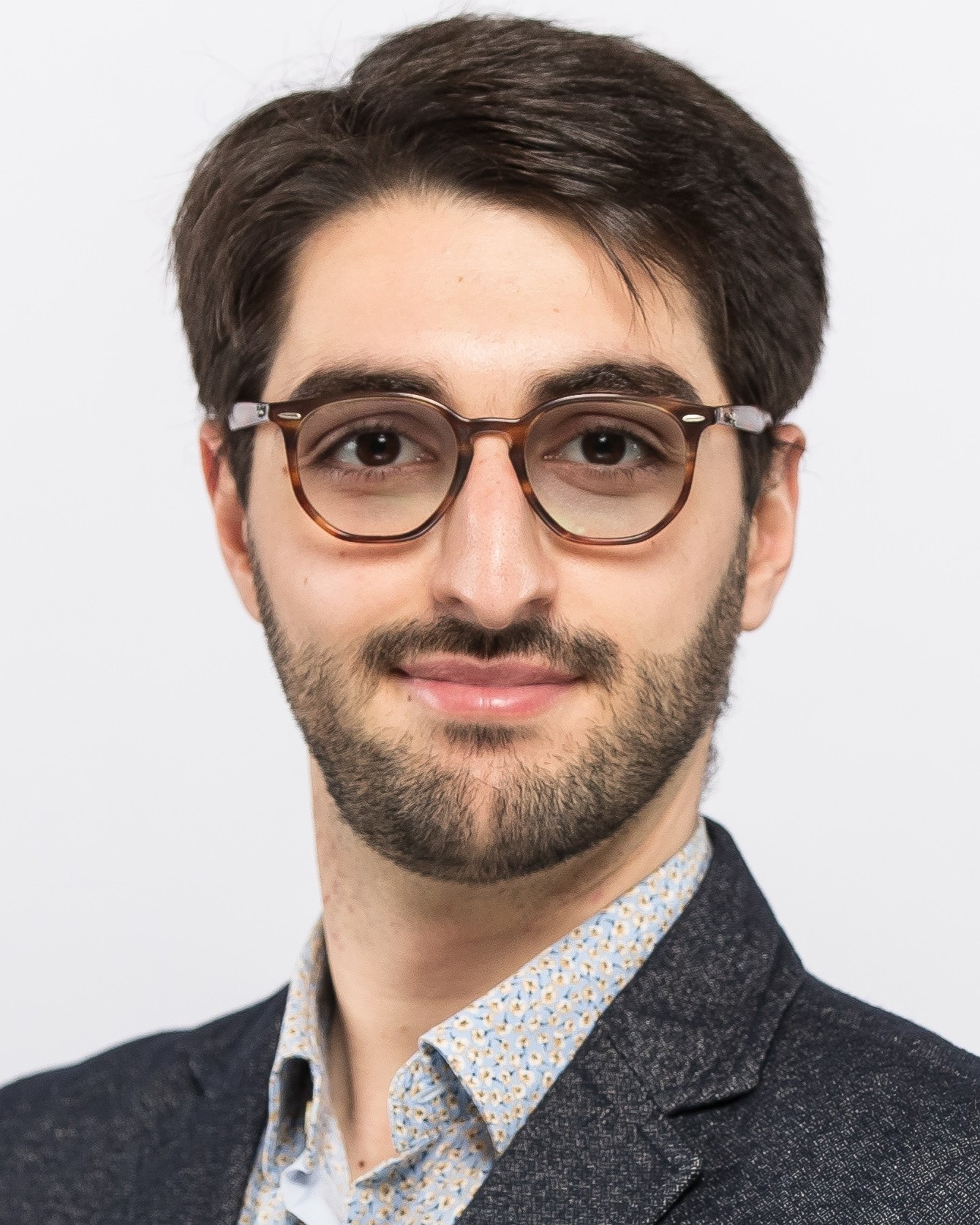}}{\textbf{Gabriele Dessena.}~\orcidlink{0000-0001-7394-9303} Gabriele has been an Assistant Professor within the Aerospace Engineering Department at the Universidad Carlos III de Madrid (Spain) since September 2023. Previously, he obtained a MEng in Aeronautics and Astronautics from the University of Southampton (UK) in 2019  and a PhD in Aerospace from Cranfield University (UK) in 2023. His research interests are centred around structural dynamics; particularly, its applications to aeronautical and aerospace structures for structural health monitoring, modal identification and nonlinearity detection. Further topics of interest are surrogate-based optimisation, mainly for the scope of finite element model updating, and unmanned aircraft and microsatellites design and analysis.}
\end{biography}

\begin{biography}{\includegraphics[width=66pt,height=86pt]{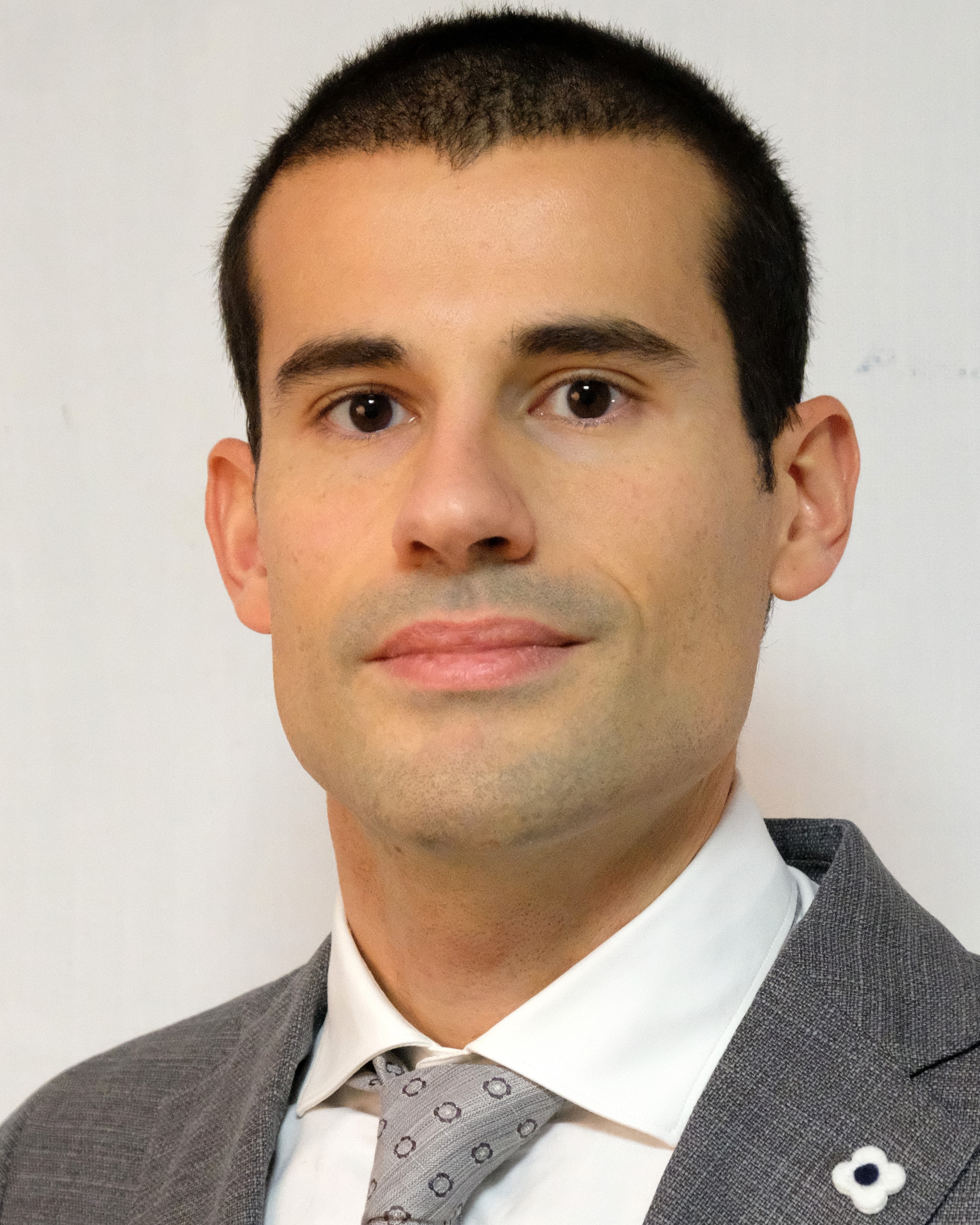}}{\textbf{Marco Civera.}~\orcidlink{0000-0003-0414-7440} Dr. Marco Civera is a Fixed-term Assistant Professor at the Department of Structural, Geotechnical, and Building Engineering (DISEG) at Politecnico di Torino. He pursued both his undergraduate and master's degrees in Civil Engineering at Politecnico di Torino, complementing his studies with experiences at École Polytechnique Fédérale de Lausanne (EPFL) and Cranfield University. He earned a Ph.D. in Aerospace Engineering from Politecnico di Torino, focusing his thesis on the vibrational response of linear and nonlinear structures. His current research interests involve Structural Dynamics, Machine Learning for Damage Detection, Nonlinear Dynamics, and Infrastructure Monitoring and Management.}
\end{biography}

\begin{biography}{\includegraphics[width=66pt,height=86pt]{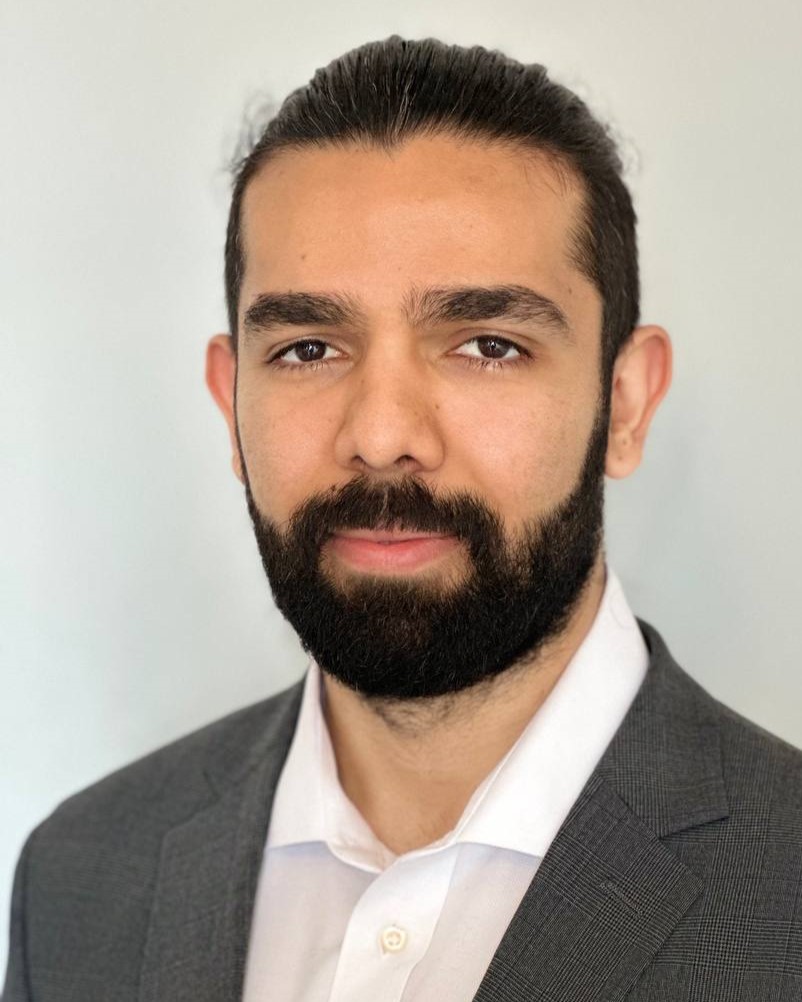}}{\textbf{Ali Yousefi.} Ali recently graduated with a Master’s degree in Civil Engineering from Politecnico di Torino, where his thesis focused on output-only system identification. He obtained his undergraduate degree from Shiraz University in 2022. Currently, he is part of the Research and Development team at Masera Engineering Group Srl, specializing in bridge optimisation design, model updating, and machine learning.
\vspace{1cm}
\phantom{This text will be invisible for formatting only.}}
\end{biography}

\begin{biography}{\includegraphics[width=66pt,height=86pt]{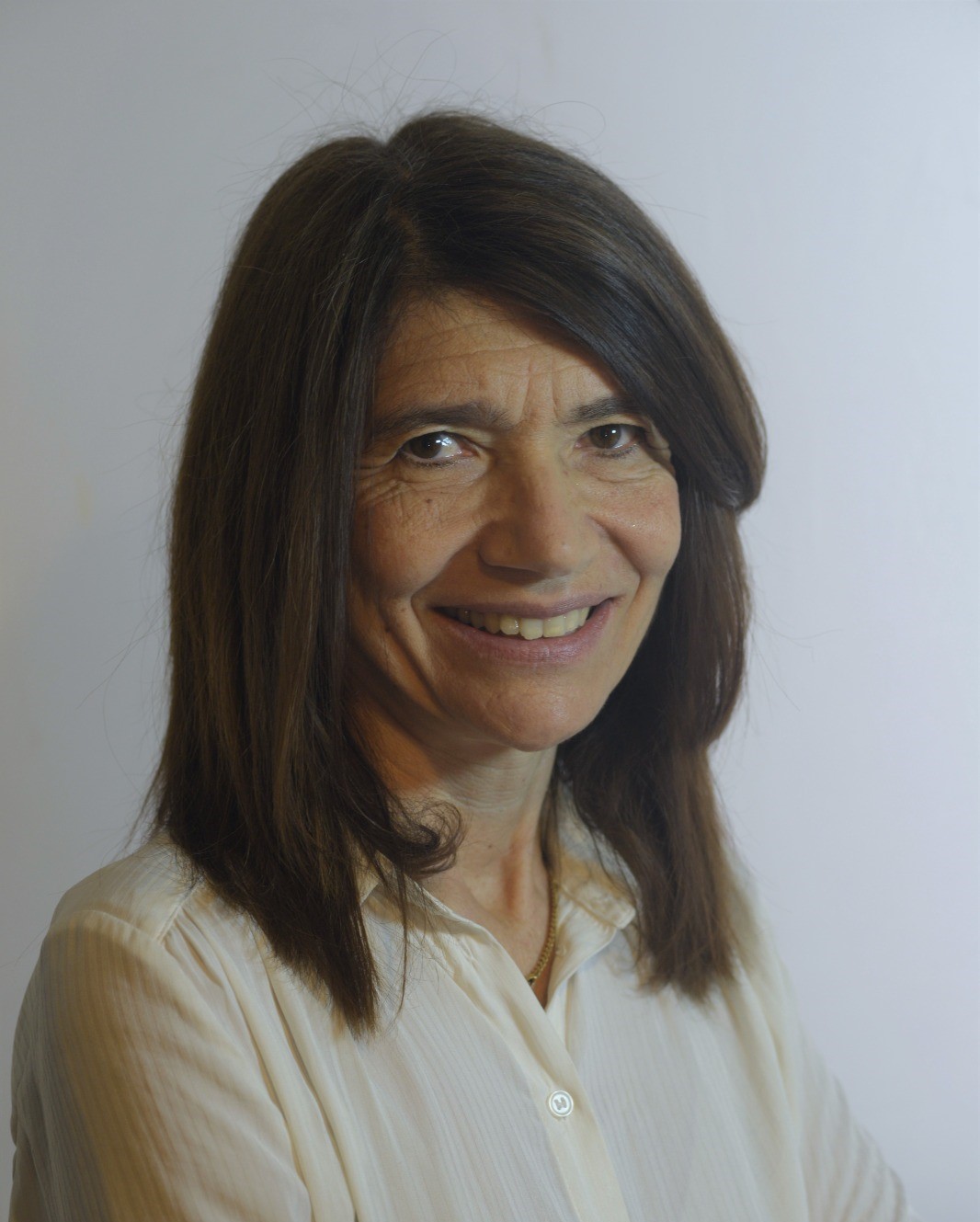}}{\textbf{Cecilia Surace.}~\orcidlink{0000-0002-3993-9432}  Dr Cecilia Surace is a Professor in Structural Mechanics at the Department of Structural, Geotechnical, and Building Engineering (DISEG) at Politecnico di Torino. With a Ph.D. in Structural Mechanics, her research interests include Dynamics of Structures, Vibration-Based Structural Health Monitoring, and Biomechanics. She is a board member of the PhD school in Aerospace Engineering at Politecnico di Torino and leads the Bio-materials and Structures (BIOMAST) laboratory at the same institute, holding several patents for medical devices and related applicator tools.}
\end{biography}

\end{document}